\documentclass[pra,preprint,superscriptaddress,floatfix,showpacs]{revtex4-1}

\usepackage{amsmath,amsfonts,amssymb}
\usepackage{graphicx}

\def\beq{\begin{equation}}
\def\eeq{\end{equation}}
\def\beqn{\begin{eqnarray}}
\def\eeqn{\end{eqnarray}}

\def\x {{\bf x}}
\def\y {{\bf y}}
\def\z {{\bf z}}

\def\Q {{\bf Q}}

\def\X {{\bf X}}
\def\Y {{\bf Y}}
\def\Z {{\bf Z}}

\begin{document}

\title{Fragmentation of a trapped multiple-species bosonic mixture}

\author{O. E. Alon}\email{ofir@research.haifa.ac.il}
\affiliation{Department of Physics, University of Haifa, 3498838 Haifa, Israel}
\affiliation{Haifa Research Center for Theoretical Physics and Astrophysics, University of Haifa, 3498838 Haifa, Israel}
\author{L. S. Cederbaum}
\affiliation{Theoretical Chemistry, Physical Chemistry Institute, Heidelberg University, D-69120 Heidelberg, Germany}

\begin{abstract}
We consider a multiple-species mixture of interacting bosons,
$N_1$ bosons of mass $m_1$,
$N_2$ bosons of mass $m_2$,
and
$N_3$ bosons of mass $m_3$
in a harmonic trap of frequency $\omega$.
The corresponding intraspecies interaction strengths are $\lambda_{11}$, $\lambda_{22}$, and $\lambda_{33}$,
and the interspecies interaction strengths are $\lambda_{12}$, $\lambda_{13}$, and $\lambda_{23}$.
When the shape of all interactions are harmonic,
this is the generic multiple-species harmonic-interaction model
which is exactly solvable.
We start by solving
the many-particle Hamiltonian and concisely discussing the ground-state
wavefunction and energy in explicit forms as functions of all parameters,
the masses, numbers of particles, and the intraspecies and interspecies interaction strengths.
We then
move to compute explicitly the reduced one-particle density matrices for all the species
and diagonalize them,
thus generalizing the treatment in [J. Chem. Phys. {\bf 161}, 184307 (2024)].
The respective eigenvalues determine the degree of fragmentation of each species.
As applications,
we focus on aspects that do not appear for the respective single-species and two-species systems.
For instance, placing a mixture of two kinds of bosons in a bath made by a third kind,
and controlling the fragmentation of the former by coupling to the latter.
Another example exploits the possibility of different connectivities (i.e., which species interacts with which species) in the mixture,
and demonstrates how the fragmentation of species $3$ can be manipulated by the interaction between species $1$ and species $2$,
when species $3$ and $1$ do not interact with each other.
We thereby highlight properties of 
fragmentation that only appear
in the multiple-species mixture.
Further applications are briefly discussed.
\end{abstract}

\maketitle 

\section{Introduction}\label{INTRO}

Fragmentation of Bose-Einstein condensates has drawn much attention for many years \cite{fr1,fr2,fr3,fr4,fr5,fr6,fr7,fr8,fr9,fr10,fr11,fr12,fr13,fr14,fr15,fr16,fr17,fr18,fr19,fr20,fr21,fr22}.
Herein, fragmentation of bosons means the macroscopic occupation of more than a single eigenvalue
of their reduced one-particle density matrix \cite{LO,YU}.
Generally, the more complex the underlying bosonic system is the more intricate its fragmentation can be.
When dealing with bosonic mixtures,
a natural question is how interspecies interactions
govern the fragmentation of individual species.
In multiple-species mixtures,
this latter question is further enriched,
which is the topic of the present work.

Multiple-species bosonic mixtures are gaining increased interest 
\cite{mm1,mm2,mm3,mm4,mm5,mm6,mm7,mm8,mm9,mm10,JCP_2024,mm12,mm13}.
Yet, treating them numerically at the many-body level of theory
is promptly becoming a difficult task,
especially when the numbers of particles of each species 
are enlarged or the interactions between particles become stronger.
To surpass this difficulty,
we refer to a solvable many-body model,
which would allow us to express analytically the reduced one-particle density matrices
and their eigenvalues as a function of the masses, numbers of bosons,
and various interactions in a generic multiple-species bosonic mixture.
Harmonic-interaction models have amply been used in the literature,
covering various setups comprising distinguishable and indistinguishable particles,
see, e.g., \cite{hm1,hm2,Robinson_1977,hm4,hm5,HIM_Cohen,HIM_MIX_IJQC,hm8,hm9,hm10,hm11,hm12,hm13,Schilling_2013,EPJD,hm16,
hm17,hm18,HIM_MIX_RDM,Floquet_HIM,Atoms_2021,BB_2022,BB_2023,BF_2024,HIM_BEC_CAVITY,FF_JPA,hm27}.
In this respect,
the present work treats the generic multiple-species harmonic-interaction model
for an imbalanced bosonic mixture.
We thereby add a new facet to the family
of exactly-solvable many-particle models in non-homogeneous environments,
i.e., in traps. 

The structure of the paper is as follows.
In Sec.~\ref{THEORY} we present the theoretical framework,
in particular the many-body Hamiltonian is diagonalized
and the reduced one-particle density matrices of the different species are evaluated explicitly,
thus generalizing the study in \cite{JCP_2024}.
In Sec.~\ref{EXAMPLES}, illustrative examples are worked out,
demonstrating intricate control of fragmentations in multiple-species mixtures. 
In Sec.~\ref{SUMMARY}, summary and outlook are put forward.
Finally, the two appendices collect complimentary information,
with appendix \ref{APP_LIMITS} discussing limiting cases of the frequencies' matrix,
associated with the centers-of-mass degrees-of-freedom,
and appendix \ref{MF} presenting a concise account of the mean-field solution
of the multiple-species bosonic mixture.

\section{Theoretical framework}\label{THEORY}

We treat an imbalanced mixture of three distinct bosonic species,
i.e., the smallest generic multiple-species mixture.
The Hamiltonian is given by
\beqn\label{HAM}
& & \hat H(\x_1,\ldots,\x_{N_1},\y_1,\ldots,\y_{N_2},\z_1,\ldots,\z_{N_3}) =
\sum_{j=1}^{N_1} \left( -\frac{1}{2m_1} \frac{\partial^2}{\partial \x_j^2} + \frac{1}{2} m_1\omega^2 \x_j^2 \right) +
\nonumber \\
& & + 
\sum_{j=1}^{N_2} \left( -\frac{1}{2m_2} \frac{\partial^2}{\partial \y_j^2} + \frac{1}{2} m_2\omega^2 \y_j^2 \right) + 
\sum_{j=1}^{N_3} \left( -\frac{1}{2m_3} \frac{\partial^2}{\partial \z_j^2} + \frac{1}{2} m_3\omega^2 \z_j^2 \right) +
\nonumber \\
& & + \lambda_1 \sum_{1 \le j < k}^{N_1} (\x_j-\x_k)^2 + 
\lambda_2 \sum_{1 \le j < k}^{N_2} (\y_j-\y_k)^2 + 
\lambda_3 \sum_{1 \le j < k}^{N_3} (\z_j-\z_k)^2 + \nonumber \\
& &
+ \lambda_{12} \sum_{j=1}^{N_1} \sum_{k=1}^{N_2} (\x_j-\y_k)^2
+ \lambda_{13} \sum_{j=1}^{N_1} \sum_{k=1}^{N_3} (\x_j-\z_k)^2
+ \lambda_{23} \sum_{j=1}^{N_2} \sum_{k=1}^{N_3} (\y_j-\z_k)^2. \
\eeqn
Here, $N_1$, $N_2$, and $N_3$ are the numbers of bosons of each species,
$m_1$, $m_2$, $m_3$ their masses,
$\lambda_1$, $\lambda_2$, and $\lambda_3$ their intraspecies interaction strengths,
and
$\lambda_{12}$, $\lambda_{13}$, and $\lambda_{23}$ are the respective
interspecies interaction strengths.
We set $\hbar=1$.
It is instructive to define the interaction parameters
$\Lambda_1=\lambda_1(N_1-1)$, $\Lambda_2=\lambda_2(N_2-1)$, and $\Lambda_3=\lambda_3(N_3-1)$,
$\Lambda_{12}=\lambda_{12}N_1$, $\Lambda_{21}=\lambda_{12}N_2$,
$\Lambda_{13}=\lambda_{13}N_1$, $\Lambda_{31}=\lambda_{13}N_3$,
$\Lambda_{23}=\lambda_{23}N_2$, and $\Lambda_{32}=\lambda_{23}N_3$,
that will be used from now on.
Finally, all bosons are trapped in a three-dimensional isotropic harmonic potential of frequency $\omega$.

To diagonalize the Hamiltonian (\ref{HAM}),
we employ a set of Jacoby coordinates for each of the species,
\beqn\label{JAC}
& & \X_s = \frac{1}{\sqrt{s(s+1)}} \sum_{j=1}^{s} (\x_{s+1}-\x_j), \ \ 1 \le s \le N_1-1, \quad \X_{N_1} = \frac{1}{\sqrt{N_1}} \sum_{j=1}^{N_1} \x_j,  \nonumber \\
& & \Y_s = \frac{1}{\sqrt{s(s+1)}} \sum_{j=1}^{s} (\y_{s+1}-\y_j), \ \ 1 \le s \le N_2-1, \quad \Y_{N_2} = \frac{1}{\sqrt{N_2}} \sum_{j=1}^{N_2} \y_j,
\nonumber \\
& & \Z_s = \frac{1}{\sqrt{s(s+1)}} \sum_{j=1}^{s} (\z_{s+1}-\z_j), \ \ 1 \le s \le N_3-1, \quad \Z_{N_3} = \frac{1}{\sqrt{N_3}} \sum_{j=1}^{N_3} \z_j. \
\eeqn
The first set of coordinates in each line is referred to as the relative-motion Jacoby coordinates
and the last coordinate in each line is called the center-of-mass Jacoby coordinate.
Transforming to the Jacoby coordinates and later on back to the laboratory frame is needed
for the various steps of evaluating quantities.

It is instrumental to define the geometric-mean mass $m_g = \left(m_1m_2m_3\right)^{\frac{1}{3}}$,
and thereafter to rescale the center-of-mass Jacoby coordinates
$\widetilde \X_{N_1} = \sqrt{\frac{m_1}{m_g}}\X_{N_1}$,
$\widetilde \Y_{N_2} = \sqrt{\frac{m_2}{m_g}}\Y_{N_2}$,
and
$\widetilde \Z_{N_3} = \sqrt{\frac{m_3}{m_g}}\Z_{N_3}$.
This transformation obviously obeys
$\widetilde \X_{N_1}\widetilde \Y_{N_2}\widetilde \Z_{N_3}=\X_{N_1}\Y_{N_2}\Z_{N_3}$. 

Then,
the Hamiltonian may be written as a sum of two Hamiltonians,
\beqn\label{HAM_MIX_3_XYZ}
& & \hat H = \hat H_{rels} + \hat H_{CMs}. \
\eeqn
The Hamiltonian of the relative motions is
\beqn\label{HAM_MIX_3_XYZ_rels}
& &
\hat H_{rels}(\X_1,\ldots,\X_{N_1-1},\Y_1,\ldots,\Y_{N_2-1},\Z_1,\ldots,\Z_{N_3-1}) =
\sum_{j=1}^{N_1-1} \left( -\frac{1}{2m_1} \frac{\partial^2}{\partial \X_j^2} + \frac{1}{2} m_1\Omega_1^2 \X_j^2 \right) +
\nonumber \\
& & + \sum_{j=1}^{N_2-1} \left( -\frac{1}{2m_2} \frac{\partial^2}{\partial \Y_j^2} + \frac{1}{2} m_2\Omega_2^2 \Y_j^2 \right)
+ \sum_{j=1}^{N_3-1} \left( -\frac{1}{2m_3} \frac{\partial^2}{\partial \Z_j^2} + \frac{1}{2} m_3\Omega_3^2 \Z_j^2 \right), \
\eeqn
where the intraspecies relative-motion frequencies are
\beqn\label{REL_OMG}
& & \Omega_1 =
\sqrt{\omega^2+\frac{2}{m_1}\left(\frac{N_1}{N_1-1}\Lambda_1+\Lambda_{21}+\Lambda_{31}\right)}, \quad
\Omega_2 = \sqrt{\omega^2+\frac{2}{m_2}\left(\frac{N_2}{N_2-1}\Lambda_2+\Lambda_{12}+\Lambda_{32}\right)}, \nonumber \\
& & \Omega_3 = \sqrt{\omega^2+\frac{2}{m_3}\left(\frac{N_3}{N_3-1}\Lambda_3+\Lambda_{13}+\Lambda_{23}\right)}. \
\eeqn
We see that, e.g., the frequency $\Omega_1$ depends on the
interactions of species $1$ with species $2$ and $3$,
but not on the interspecies interaction between the latter two.
We shall return to this issue below, also see appendix \ref{MF}.

The center-of-masses Hamiltonian reads
\beqn\label{HAM_MIX_3_XYZ_CMs}
& &
\hat H_{CMs}(\widetilde\X_{N_1},\widetilde\Y_{N_2},\widetilde\Z_{N_3}) = \nonumber \\
& &
= - \frac{1}{2m_g} \left(\frac{\partial^2}{\partial \widetilde\X_{N_1}^2} + \frac{\partial^2}{\partial \widetilde\Y_{N_2}^2}
+ \frac{\partial^2}{\partial \widetilde\Z_{N_3}^2}\right)
+ \frac{1}{2} m_g
\begin{pmatrix}
\widetilde\X_{N_1} & \widetilde\Y_{N_2} & \widetilde\Z_{N_3} \cr
\end{pmatrix}
\underline{\underline{\bf O}}
\begin{pmatrix}
\widetilde\X_{N_1} \cr \widetilde\Y_{N_2} \cr \widetilde\Z_{N_3} \cr
\end{pmatrix}, \nonumber \\
& &
\underline{\underline{\bf O}} =
\begin{pmatrix}
\omega^2 + \frac{2}{m_1}(\Lambda_{21}+\Lambda_{31})
& -\frac{2}{m_1}\Lambda_{21}\sqrt{\frac{m_1N_1}{m_2N_2}} 
& -\frac{2}{m_1}\Lambda_{31}\sqrt{\frac{m_1N_1}{m_3N_3}} \cr
-\frac{2}{m_2}\Lambda_{12}\sqrt{\frac{m_2N_2}{m_1N_1}} 
& \omega^2 + \frac{2}{m_2}(\Lambda_{12}+\Lambda_{32}) 
& -\frac{2}{m_2}\Lambda_{32}\sqrt{\frac{m_2N_2}{m_3N_3}}  \cr
-\frac{2}{m_3}\Lambda_{13}\sqrt{\frac{m_3N_3}{m_1N_1}}
& -\frac{2}{m_3}\Lambda_{23}\sqrt{\frac{m_3N_3}{m_2N_2}}
& \omega^2 + \frac{2}{m_3}(\Lambda_{13}+\Lambda_{23}) \cr
\end{pmatrix}. \
\eeqn
The frequencies' matrix $\underline{\underline{\bf O}}$
governs the coupling of the center-of-mass coordinates and
is expressed as a function of
all interspecies interaction parameters,
$\Lambda_{12}$, $\Lambda_{21}$, $\Lambda_{13}$, $\Lambda_{31}$, $\Lambda_{23}$, and $\Lambda_{32}$,
for convenience.
It is of course symmetric.

Diagonalizing $\underline{\underline{\bf O}}$,
the frequencies of the center-of-masses Hamiltonian are
\beqn\label{CM_OMG}
& & \Omega_{123}^\pm = \sqrt{\omega^2 + K \pm \sqrt{K^2 - 4G}}, \quad
\omega, \
\eeqn
where
\beqn\label{CM_3_freq_KG}
& & K=\frac{1}{m_1}\left(\Lambda_{21}+\Lambda_{31}\right) + \frac{1}{m_2}\left(\Lambda_{12}+\Lambda_{32}\right)
+ \frac{1}{m_3}\left(\Lambda_{13}+\Lambda_{23}\right), \nonumber \\
& & G=
\frac{1}{m_1m_2}\left(\Lambda_{12}\Lambda_{31}+\Lambda_{31}\Lambda_{32}+\Lambda_{32}\Lambda_{21}\right)+
\frac{1}{m_1m_3}\left(\Lambda_{13}\Lambda_{21}+\Lambda_{21}\Lambda_{23}+\Lambda_{23}\Lambda_{31}\right)+ \nonumber \\
& & +
\frac{1}{m_2m_3}\left(\Lambda_{23}\Lambda_{12}+\Lambda_{12}\Lambda_{13}+\Lambda_{13}\Lambda_{32}\right). \
\eeqn
The structure of the frequencies $\Omega_{123}^\pm$ is worth a discussion.
$K$ is referred to as the two-body part and $G$ the three-body part which naturally cannot exist in a mixture with two species.
We shall look into
the impact of the latter below.
Finally, as all species are trapped in the same harmonic potential,
the center-of-mass frequency is $\omega$.

Summing up all together,
the ground-state energy of the generic three-species mixture
is given by
\beqn\label{E_GEN}
& & E = E_{rels} + E_{CMs}, \
\eeqn
where
\beqn\label{E_GEN_RELS}
&&
E_{rels} = \frac{3}{2} \left[ (N_1-1) \Omega_1 + (N_2-1) \Omega_2 + (N_3-1) \Omega_3 \right] = \nonumber \\
& &
= \frac{3}{2}
\Bigg[(N_1-1) \sqrt{\omega^2+\frac{2}{m_1}\left(\frac{N_1}{N_1-1}\Lambda_1+\Lambda_{21}+\Lambda_{31}\right)} + \nonumber \\
& &
+ (N_2-1)\sqrt{\omega^2+\frac{2}{m_2}\left(\frac{N_2}{N_2-1}\Lambda_2+\Lambda_{12}+\Lambda_{32}\right)} + \nonumber \\
& &
+ (N_3-1)\sqrt{\omega^2+\frac{2}{m_3}\left(\frac{N_3}{N_3-1}\Lambda_3+\Lambda_{13}+\Lambda_{23}\right)}\Bigg] \
\eeqn
and
\beqn\label{E_GEN_CMs}
& & 
E_{CMs} =  \frac{3}{2} \left[ \Omega_{123}^+ + \Omega_{123}^- + \omega \right] = \nonumber \\
& &
= \frac{3}{2} \left[ \sqrt{\omega^2 + K  + \sqrt{K^2 - 4G}} + \sqrt{\omega^2 + K  - \sqrt{K^2 - 4G}} + \omega \right]. \
\eeqn 
Of course, the frequencies $\Omega_1$, $\Omega_2$, $\Omega_3$, $\Omega_{123}^+$, and $\Omega_{123}^-$
must all be positive for the three-species mixture to be bound.
This sets five simultaneous restrictions on the interaction parameters,
\beqn\label{FREQ_COND}
& &
\frac{N_1}{N_1-1}\Lambda_1+\Lambda_{21}+\Lambda_{31} > - \frac{m_1\omega^2}{2}, \quad
\frac{N_2}{N_2-1}\Lambda_2+\Lambda_{12}+\Lambda_{32} > - \frac{m_2\omega^2}{2}, \nonumber \\
& &
\frac{N_3}{N_3-1}\Lambda_3+\Lambda_{13}+\Lambda_{23} > - \frac{m_3\omega^2}{2}, \quad
K \pm \sqrt{K^2 - 4G} > - \omega^2. \
\eeqn
As a result, the energy (\ref{E_GEN}) is bound from below by the frequency of the trap
$\omega$ but is not bound from above.
Physically,
the lower bounds (\ref{FREQ_COND}) mean that the overall repulsion between the bosons
cannot be too strong.
In Sec.~\ref{EXAMPLES}, we present an example and emerging effects at
the edge of stability of the mixture.

The ground-state wavefunction then takes on the separable form
\beqn\label{WF_HIM_3}
& & \Psi(\X_1,\ldots,\Y_1,\ldots,\Z_1,\ldots,\Q_1,\Q_2,\Q_3) = \nonumber \\
& & =
\left(\frac{m_1\Omega_1}{\pi}\right)^{\frac{3(N_1-1)}{4}}
\left(\frac{m_2\Omega_2}{\pi}\right)^{\frac{3(N_2-1)}{4}}
\left(\frac{m_3\Omega_3}{\pi}\right)^{\frac{3(N_3-1)}{4}}
\left(\frac{m_g\Omega^+_{123}}{\pi}\right)^{\frac{3}{4}}
\left(\frac{m_g\Omega^-_{123}}{\pi}\right)^{\frac{3}{4}}
\left(\frac{m_g\omega}{\pi}\right)^{\frac{3}{4}} \times \nonumber \\
& & \times
e^{-\frac{1}{2} \left(m_1\Omega_1 \sum_{k=1}^{N_1-1} \X_k^2
+ m_2\Omega_2 \sum_{k=1}^{N_2-1} \Y_k^2
+ m_3\Omega_3 \sum_{k=1}^{N_3-1} \Z_k^2\right)}
e^{-\frac{1}{2}m_g\left(\Omega^+_{123} \Q_1^2 + \Omega^-_{123} \Q_2^2 + \omega \Q_3^2\right)}, \
\eeqn
where the (orthonormal) eigenvectors of the center-of-masses Hamiltonian are denoted as
\beqn\label{CM_3_vecs}
& &
\Q_J = \Delta^{(J)}_{1} \widetilde\X_{N_1} + \Delta^{(J)}_{2} \widetilde\Y_{N_2} + \Delta^{(J)}_{3} \widetilde\Z_{N_2},
\quad 1 \le J \le 3. \
\eeqn
The explicit expressions for the $\Delta^{(J)}_{K}$ are depicted hereafter.

The components of the two relative center-of-mass coordinates $\Q_1$ and $\Q_2$
for the case of general nondegenerate roots $\Omega_{123}^\pm$ (\ref{CM_OMG}) are
\beqn\label{BBB_CM_REL}
& &
\Delta^{(1)}_{1}, \Delta^{(2)}_{1} = {\mathcal{N}}_\pm \sqrt{\frac{m_1N_1}{m_3N_3}}
\left[\frac{\Lambda_{31}}{m_1}
\left(\frac{\Lambda_{12}+\Lambda_{32}}{m_2}-\frac{K \pm \sqrt{K^2-4G}}{2}\right)
+ \frac{\Lambda_{21}\Lambda_{32}}{m_1m_2}\right], \nonumber \\
& &
\Delta^{(1)}_{2}, \Delta^{(2)}_{2} = {\mathcal{N}}_\pm \sqrt{\frac{m_2N_2}{m_3N_3}}
\left[\frac{\Lambda_{32}}{m_2}
\left(\frac{\Lambda_{21}+\Lambda_{31}}{m_1}-\frac{K \pm \sqrt{K^2-4G}}{2}\right)
+ \frac{\Lambda_{12}\Lambda_{31}}{m_1m_2}\right], \nonumber \\
& &
\Delta^{(1)}_{3}, \Delta^{(2)}_{3} = \\
& &
= {\mathcal{N}}_\pm
\left[
\left(\frac{\Lambda_{12}+\Lambda_{32}}{m_2}-
\frac{K \pm \sqrt{K^2-4G}}{2}\right)
\left(\frac{\Lambda_{21}+\Lambda_{31}}{m_1}-
\frac{K \pm \sqrt{K^2-4G}}{2}\right)
- \frac{\Lambda_{12}\Lambda_{21}}{m_1m_2}
\right], \nonumber \
\eeqn
where ${\mathcal{N}}_\pm$ are the normalizations.
Clearly, the components of the relative center-of-mass coordinates depend
on the interspecies interaction strengths,
unlike the corresponding relative coordinate of the two-species mixture or the
relative coordinates in the specific case
of a balanced multiple-species mixture \cite{JCP_2024}.

In all cases of a three-species mixture,
the components of the center-of-mass coordinate $\Q_3$ are, of course,
\beqn\label{BBB_CM_ALL}
& &
\Delta^{(3)}_{1} =
\sqrt{\frac{m_1N_1}{m_1N_1+m_2N_2+m_3N_3}}, \quad
\Delta^{(3)}_{2} =
\sqrt{\frac{m_2N_2}{m_1N_1+m_2N_2+m_3N_3}}, \nonumber \\
& &
\Delta^{(3)}_{3} =
\sqrt{\frac{m_3N_3}{m_1N_1+m_2N_2+m_3N_3}}, \
\eeqn
and do not depend on the interaction strengths.

In using the general expressions (\ref{BBB_CM_REL}) and (\ref{BBB_CM_ALL}) explicitly,
specific cases and limits should be noted and discussed separately.
For the ease of reading,
these cases are collected and analyzed
in appendix \ref{APP_LIMITS}

To proceed and calculate explicitly the reduced one-particle density matrices
of species $1$, $2$, and $3$
we work in the representation of the wavefunction using the
Jacoby coordinates of each species explicitly,
\beqn\label{WF_HIM_3_JAC}
& & \Psi(\X_1,\ldots,\X_{N_1},\Y_1,\ldots,\Y_{N_2},\Z_1,\ldots,\Z_{N_3}) = \nonumber \\
& & = \left(\frac{m_1\Omega_1}{\pi}\right)^{\frac{3(N_1-1)}{4}}
\left(\frac{m_2\Omega_2}{\pi}\right)^{\frac{3(N_2-1)}{4}}
\left(\frac{m_3\Omega_3}{\pi}\right)^{\frac{3(N_3-1)}{4}}
\left(\frac{m_g\Omega^+_{123}}{\pi}\right)^{\frac{3}{4}}
\left(\frac{m_g\Omega^-_{123}}{\pi}\right)^{\frac{3}{4}}
\left(\frac{m_g\omega}{\pi}\right)^{\frac{3}{4}} \times \nonumber \\
& & \times
e^{-\frac{1}{2} m_1\Omega_1 \sum_{k=1}^{N_1-1} \X_k^2}
e^{-\frac{1}{2} m_2\Omega_2 \sum_{k=1}^{N_2-1} \Y_k^2}
e^{-\frac{1}{2} m_3\Omega_3 \sum_{k=1}^{N_3-1} \Z_k^2}
e^{-\frac{1}{2} a_1 \X_{N_1}^2}
e^{-\frac{1}{2} a_2 \Y_{N_2}^2}
e^{-\frac{1}{2} a_2 \Z_{N_3}^2} \times \nonumber \\
& &
\times
e^{-b_{12} \X_{N_1}\Y_{N_2}}
e^{-b_{13} \X_{N_1}\Z_{N_3}}
e^{-b_{23} \Y_{N_2}\Z_{N_3}}, \
\eeqn
where, to recall, $m_g = \left(m_1m_2m_3\right)^{\frac{1}{3}}$.
The various coefficients of the
center-of-masses part are
\beqn\label{WF_HIM_3_JACS_COEFF}
& &
\!\!\!\!\!\!\!\!\!\!\!\!
a_J = m_J\left(\left\{\Delta^{(1)}_{J}\right\}^2\Omega^+_{123}
+ \left\{\Delta^{(2)}_{J}\right\}^2\Omega^-_{123}
+ \left\{\Delta^{(3)}_{J}\right\}^2\omega\right), \quad 1 \le J \le 3,  \nonumber \\
& &
\!\!\!\!\!\!\!\!\!\!\!\!
b_{JK} = \sqrt{m_Jm_K}\left(\Delta^{(1)}_{J}\Delta^{(1)}_{K}\Omega^+_{123}
+\Delta^{(2)}_{J}\Delta^{(2)}_{K}\Omega^-_{123}
+\Delta^{(3)}_{J}\Delta^{(3)}_{K}\omega\right), \quad 1\le J<K \le 3. \
\eeqn
Then, the all-particle density matrix expressed using the species' Jacoby coordinates is just
\beqn\label{N1_N2_N3_DENS_JAC}
& & \Psi(\X_1,\ldots,\X_{N_1},\Y_1,\ldots,\Y_{N_2},\Z_1,\ldots,\Z_{N_3})
\Psi^\ast(\X'_1,\ldots,\X'_{N_1},\Y'_1,\ldots,\Y'_{N_2},\Z'_1,\ldots,\Z'_{N_3}) = \nonumber \\
& & =
\left(\frac{m_1\Omega_1}{\pi}\right)^{\frac{3(N_1-1)}{2}}
\left(\frac{m_2\Omega_2}{\pi}\right)^{\frac{3(N_2-1)}{2}}
\left(\frac{m_3\Omega_3}{\pi}\right)^{\frac{3(N_3-1)}{2}}
\left(\frac{m_g\Omega^+_{123}}{\pi}\right)^{\frac{3}{2}}
\left(\frac{m_g\Omega^-_{123}}{\pi}\right)^{\frac{3}{2}}
\left(\frac{m_g\omega}{\pi}\right)^{\frac{3}{2}} \times \nonumber \\
& & \times
e^{-\frac{1}{2} m_1\Omega_1 \sum_{k=1}^{N_1-1} \left(\X_k^2+{\X'}_k^2\right)}
e^{-\frac{1}{2} m_2\Omega_2 \sum_{k=1}^{N_2-1} \left(\Y_k^2+{\Y'}_k^2\right)}
e^{-\frac{1}{2} m_3\Omega_3 \sum_{k=1}^{N_3-1} \left(\Z_k^2+{\Z'}_k^2\right)}
\times \nonumber \\
& & 
\times
e^{-\frac{1}{2} a_1 \left(\X_{N_1}^2+{\X'}_{N_1}^2\right)}
e^{-\frac{1}{2} a_2 \left(\Y_{N_2}^2+{\Y'}_{N_2}^2\right)}
e^{-\frac{1}{2} a_3 \left(\Z_{N_3}^2+{\Z'}_{N_3}^2\right)}
e^{-b_{12} \left(\X_{N_1}\Y_{N_2}+\X'_{N_1}\Y'_{N_2}\right)}
\times \nonumber \\
& &
\times
e^{-b_{13} \left(\X_{N_1}\Z_{N_3}+\X'_{N_1}\Z'_{N_3}\right)}
e^{-b_{23}\left(\Y_{N_2}\Z_{N_3}+\Y'_{N_2}\Z'_{N_3}\right)}, \
\eeqn
where normalization to unity is employed.
The all-particle density matrix (\ref{N1_N2_N3_DENS_JAC}) 
contains the correlations in the ground state to all orders,
and contracting it by integrating selected degrees-of-freedom
expresses these correlations via reduced density matrices.

The integration scheme for the one-particle reduced density matrix of species $1$
starts by eliminating of the relative-motion Jacoby coordinates of species $3$ and $2$
and proceeds
over the center-of-mass of species $3$, $\Z'_{N_3}=\Z_{N_3}$, and that of species $2$, $\Y'_{N_2}=\Y_{N_2}$,
and gives
\beqn\label{INT_CM_BBB_3}
& & \int d\Y_{N_2} e^{-a_2 \Y_{N_2}^2}
e^{-b_{12}\left(\X_{N_1}+\X'_{N_1}\right)\Y_{N_2}}
\int d\Z_{N_3} e^{-a_3 \Z_{N_3}^2}
e^{-\left[b_{13}\left(\X_{N_1}+\X'_{N_1}\right)+2b_{23}\Y_{N_2}\right]\Z_{N_3}}
= \nonumber \\
& &
= \left(\frac{\pi^2}{a_2a_3-b_{23}^2}\right)^{\frac{3}{2}}
e^{+\frac{1}{4}\left(\frac{a_2b_{13}^2+a_3b_{12}^2-2b_{12}b_{13}b_{23}}{a_2a_3-b_{23}^2}\right)\left(\X_{N_1}+\X'_{N_1}\right)^2}. \
\eeqn
Hence, we see how the center-of-mass of the remaining species $1$
gets dressed by the centers-of-mass of the others species.
Expression (\ref{INT_CM_BBB_3}) generalizes
the respective one
in \cite{JCP_2024} obtained for the balanced three-species mixture.
Further, by performing the last integration, 
over the center-of-mass of species $1$, $\X'_{N_1}=\X_{N_1}$,
keeping in mind that the normalizations of (\ref{WF_HIM_3}) and (\ref{WF_HIM_3_JAC}) are alike,
we arrive at a useful relation between all coefficients,
\beqn\label{THE_As_and_Bs}
a_1a_2a_3+2b_{12}b_{13}b_{23}-\left(a_1b_{23}^2+a_2b_{13}^2+a_3b_{12}^2\right)=m_1m_2m_3\Omega^+_{123}\Omega^-_{123}\omega.
\eeqn
This relation is instrumental in simplifying the reduced one-particle density matrices, see below.
  
The
resulting expression for the $N_1$-particle reduced density matrix of species $1$
in the three-species mixture is given by
\beqn\label{F_1_0_0_BBB_3}
& &
\rho^{(N_1)}_1(\x_1,\ldots,\x_{N_1},\x'_1,\ldots,\x'_{N_1}) = 
N_1!
\left[\left(\frac{\alpha_1-\beta_1}{\pi}\right)^{N_1-1}
\frac{\left(\alpha_1-\beta_1\right)+N_1\left(\beta_1+C_{N_1,0,0}\right)}
{\pi}\right]^{\frac{3}{2}}
\times \nonumber \\
& &
\times e^{-\frac{\alpha_1}{2} \sum_{j=1}^{N_1} \left(\x_j^2+{\x'}_j^2\right) - \beta_1 \sum_{1\le j <k}^{N_1}
\left(\x_j \x_k + \x'_j \x'_k\right)} e^{-\frac{1}{4}C_{N_1,0,0}\left\{\sum_{j=1}^{N_1}\left(\x_j+\x'_j\right)\right\}^2}, \
\eeqn
with
\beqn\label{F_1_0_0_BBB_3_Coeff}
& & \alpha_1 =
m_1\Omega_1 + \beta_1, \qquad
\beta_1 =
\frac{1}{N_1}\left(a_1-m_1\Omega_1\right),
\nonumber \\
& & C_{N_1,0,0} = - \frac{1}{N_1}\left(\frac{a_2b_{13}^2+a_3b_{12}^2-2b_{12}b_{13}b_{23}}{a_2a_3-b_{23}^2}\right), \
\eeqn
where, to recall, 
$a_1 = m_1\left(\left\{\Delta^{(1)}_{1}\right\}^2\Omega^+_{123}
+ \left\{\Delta^{(2)}_{1}\right\}^2\Omega^-_{123}
+ \left\{\Delta^{(3)}_{1}\right\}^2\omega\right)$
and the other coefficients
$a_J$ and $b_{JK}$
are collected in (\ref{WF_HIM_3_JACS_COEFF}).

Hence, the one-particle reduced density matrix of species $1$ is integrated from (\ref{F_1_0_0_BBB_3},\ref{F_1_0_0_BBB_3_Coeff})
and given by
\beqn\label{rho_1_BBB_3}
& & \rho^{(1)}_1(\x,\x') = N_1 \left(\frac{\alpha_1+C_{1,0,0}}{\pi}\right)^{\frac{3}{2}}
e^{-\frac{\alpha_1}{2}\left(\x^2+{\x'}^2\right)}e^{-\frac{1}{4}C_{1,0,0}\left(\x+\x'\right)^2}, \nonumber \\
& &
\alpha_1+C_{1,0,0} =
\left(\alpha_1-\beta_1\right) \frac{\left(\alpha_1-\beta_1\right) + N_1\left(C_{N_1,0,0}+\beta_1\right)}
{\left(\alpha_1-\beta_1\right) + (N_1-1)\left(C_{N_1,0,0}+\beta_1\right)}. \
\eeqn
The one-particle reduced density matrix in the generic
multiple-species mixture 
(\ref{rho_1_BBB_3}) has the same structure as in simpler cases
\cite{HIM_Cohen,HIM_MIX_RDM,JCP_2024},
albeit with more complex explicit coefficients, also see below. 
The diagonal, $\rho^{(1)}_1(\x) = N_1 \left(\frac{\alpha_1+C_{1,0,0}}{\pi}\right)^{\frac{3}{2}}
e^{-\left(\alpha_1+C_{1,0,0}\right)\x^2}$, is simply the density of species $1$.
The corresponding expressions for the one-particle reduced density matrices of species $2$ and $3$ are obtained in the same way
and thus relate to (\ref{rho_1_BBB_3}) by the appropriate interchanges of quantities or indexes,
\beqn\label{rho_2_BBB_3}
& & \rho^{(1)}_2(\y,\y') = N_2 \left(\frac{\alpha_2+C_{0,1,0}}{\pi}\right)^{\frac{3}{2}}
e^{-\frac{\alpha_1}{2}\left(\y^2+{\y'}^2\right)}e^{-\frac{1}{4}C_{0,1,0}\left(\y+\y'\right)^2}, \nonumber \\
& &
\alpha_2+C_{0,1,0} =
\left(\alpha_2-\beta_2\right) \frac{\left(\alpha_2-\beta_2\right) + N_2\left(C_{0,N_2,0}+\beta_2\right)}
{\left(\alpha_2-\beta_2\right) + (N_2-1)\left(C_{0,N_2,0}+\beta_2\right)} \
\eeqn
and
\beqn\label{rho_3_BBB_3}
& & \rho^{(1)}_3(\z,\z') = N_3 \left(\frac{\alpha_3+C_{0,0,1}}{\pi}\right)^{\frac{3}{2}}
e^{-\frac{\alpha_1}{2}\left(\z^2+{\z'}^2\right)}e^{-\frac{1}{4}C_{0,0,1}\left(\z+\z'\right)^2}, \nonumber \\
& &
\alpha_3+C_{0,0,1} =
\left(\alpha_3-\beta_3\right) \frac{\left(\alpha_3-\beta_3\right) + N_3\left(C_{0,0,N_3}+\beta_3\right)}
{\left(\alpha_3-\beta_3\right) + (N_3-1)\left(C_{0,0,N_3}+\beta_3\right)}, \
\eeqn
where
$C_{0,N_2,0} = - \frac{1}{N_2}\left(\frac{a_1b_{23}^2+a_3b_{12}^2-2b_{12}b_{13}b_{23}}{a_1a_3-b_{13}^2}\right)$
and
$C_{0,0,N_3} = - \frac{1}{N_3}\left(\frac{a_1b_{23}^2+a_2b_{13}^2-2b_{12}b_{13}b_{23}}{a_1a_2-b_{12}^2}\right)$,
and
$\rho^{(1)}_2(\y) = N_2 \left(\frac{\alpha_2+C_{0,1,0}}{\pi}\right)^{\frac{3}{2}}
e^{-\left(\alpha_2+C_{0,1,0}\right)\y^2}$
and
$\rho^{(1)}_3(\z) = N_3 \left(\frac{\alpha_3+C_{0,0,1}}{\pi}\right)^{\frac{3}{2}}
e^{-\left(\alpha_3+C_{0,0,1}\right)\z^2}$
are the one-particle densities of species $2$ and $3$, respectively.
For completeness,
the resulting expressions for the $N_2$-particle reduced density matrix of species $2$
and $N_3$-particle reduced density matrix of species $3$
in the three-species mixture are given analogously to (\ref{F_1_0_0_BBB_3})
by the corresponding interchanges of indexes.

To compute the depletion of each of the species,
diagonalization of the above reduced one-particle density matrices
with Mehler's formula \cite{Robinson_1977,Schilling_2013,Atoms_2021,JCP_2024}
is performed in three spatial dimensions,
$\left[\frac{(1-\rho)s}{(1+\rho)\pi}\right]^{\frac{3}{2}}
e^{-\frac{1}{2}\frac{(1+\rho^2)s}{1-\rho^2}\left(\x^2+{\x'}^2\right)}e^{+\frac{2\rho s}{1-\rho^2}\x\x'} = 
\sum_{n_1,n_2,n_3=0}^\infty (1-\rho)^3\rho^{n_1+n_2+n_3}\break\hfill
\Phi_{n_1,n_2,n_3}(\x;s) \Phi_{n_1,n_2,n_3}(\x';s)$,
where
$\Phi_{n_1,n_2,n_3}(\x;s) =
\frac{1}{\sqrt{2^{n_1+n_2+n_3} n_1! n_2! n_3!}}
\left(\frac{s}{\pi}\right)^{\frac{3}{4}} H_{n_1}(\sqrt{s}x_1)\break\hfill H_{n_2}(\sqrt{s}x_2) H_{n_3}(\sqrt{s}x_3) e^{-\frac{1}{2}s \x^2}$.
Here, $0 < s$,
$0 \le \rho <1$,
$H_n(x)$ are the Hermite polynomials,
and $\x=(x_1,x_2,x_3)$.

Thus, the eigenvalues of each of the reduced one-particle density matrices are generated by the respective parameter
$\rho$ appearing in the Mehler's formula.
The first few occupation numbers per particle,
as they are often called, are
\beqn\label{RHO_EIGEN}
(1-\rho)^3, \quad (1-\rho)^3\rho \mathrm{\ \ (multiplicity \ 3)}, \quad (1-\rho)^3\rho^2  \mathrm{\ (multiplicity \ 6)}, \, \ldots.
\eeqn
Correspondingly,
the fraction of depleted bosons, i.e., bosons residing in all the natural orbitals but the first (lowest), is
\beqn\label{RHO_DEPLETE}
d=1-(1-\rho)^3=\rho \left(3-3\rho+\rho^2\right),
\eeqn
where $\rho$ is computed hereafter
explicitly for each of the species.
In what follows,
we investigate the fragmentation and depletion of species $1$, $2$, and $3$.
Following the relations (\ref{RHO_EIGEN}) and (\ref{RHO_DEPLETE}),
we shall use the terms fragmentation and depletion interchangeably.
For species $1$, we find explicitly
{\footnotesize{
\beqn\label{1RDM_RHO_S1}
& &
\!\!\!\!\!\!
\rho^{(1)}_1 = \frac{\mathcal{W}_1-1}{\mathcal{W}_1+1}, \\
& &
\!\!\!\!\!\!
\mathcal{W}_1 = \sqrt{\frac{\alpha_1}{\alpha_1+C_{1,0,0}}} =
\sqrt{1+\frac{1}{N_1}\left(\left\{\Delta^{(1)}_{1}\right\}^2\frac{\Omega^+_{123}}{\Omega_1}
+ \left\{\Delta^{(2)}_{1}\right\}^2\frac{\Omega^-_{123}}{\Omega_1}
+ \left\{\Delta^{(3)}_{1}\right\}^2\frac{\omega}{\Omega_1} - 1\right)} \times \nonumber \\
& &
\!\!\!\!\!\!
\sqrt{1+\frac{1}{N_1}\left(
\left\{\Delta^{(2)}_{2} \Delta^{(3)}_{3} - \Delta^{(2)}_{3} \Delta^{(3)}_{2}\right\}^2 \frac{\Omega_1}{\Omega^+_{123}} +
\left\{\Delta^{(1)}_{2} \Delta^{(3)}_{3} - \Delta^{(1)}_{3} \Delta^{(3)}_{2}\right\}^2 \frac{\Omega_1}{\Omega^-_{123}} +
\left\{\Delta^{(1)}_{2} \Delta^{(2)}_{3} - \Delta^{(1)}_{3} \Delta^{(2)}_{2}\right\}^2 \frac{\Omega_1}{\omega} - 1\right)}, \nonumber \ 
\eeqn
}}
where (\ref{THE_As_and_Bs}) is used.
Eq.~(\ref{1RDM_RHO_S1}) is one of the main results of the present work and 
describes the fragmentation of species $1$ as an explicit function of all parameters in the mixture,
the masses, numbers of particles, intraspecies and interspecies interactions in (\ref{HAM}), i.e.,
all together twelve different paramters!
The multiple-species result (\ref{1RDM_RHO_S1}) generalizes the two-species expression \cite{BB_2023}.
Conceptually, there are not only more coefficients in the former but, as we have seen above,
the relative center-of-mass coefficients (\ref{BBB_CM_REL}) are interaction-dependent,
unlike in the two-species mixture,
which can support richer fragmentation scenarios.
The respective expressions for the fragmentations
of species $2$ and $3$ are obtained analogously
and hence relate to (\ref{1RDM_RHO_S1}) by the appropriate interchanges of indexes,
{\footnotesize{
\beqn\label{1RDM_RHO_S2}
& &
\!\!\!\!
\rho^{(1)}_2 = \frac{\mathcal{W}_2-1}{\mathcal{W}_2+1}, \\
& &
\!\!\!\!
\mathcal{W}_2 = \sqrt{\frac{\alpha_2}{\alpha_2+C_{0,1,0}}} =
\sqrt{1+\frac{1}{N_2}\left(\left\{\Delta^{(1)}_{2}\right\}^2\frac{\Omega^+_{123}}{\Omega_2}
+ \left\{\Delta^{(2)}_{2}\right\}^2\frac{\Omega^-_{123}}{\Omega_2}
+ \left\{\Delta^{(3)}_{2}\right\}^2\frac{\omega}{\Omega_2} - 1\right)} \times \nonumber \\
& &
\!\!\!\!
\sqrt{1+\frac{1}{N_2}\left(
\left\{\Delta^{(2)}_{1} \Delta^{(3)}_{3} - \Delta^{(2)}_{3} \Delta^{(3)}_{1}\right\}^2 \frac{\Omega_2}{\Omega^+_{123}} +
\left\{\Delta^{(1)}_{1} \Delta^{(3)}_{3} - \Delta^{(1)}_{3} \Delta^{(3)}_{1}\right\}^2 \frac{\Omega_2}{\Omega^-_{123}} +
\left\{\Delta^{(1)}_{1} \Delta^{(2)}_{3} - \Delta^{(1)}_{3} \Delta^{(2)}_{1}\right\}^2 \frac{\Omega_2}{\omega} - 1\right)} \nonumber \
\eeqn
}}
and
{\footnotesize{
\beqn\label{1RDM_RHO_S3}
& &
\!\!\!\!\!\!
\rho^{(1)}_3 = \frac{\mathcal{W}_3-1}{\mathcal{W}_3+1}, \\
& &
\!\!\!\!\!\!
\mathcal{W}_3 = \sqrt{\frac{\alpha_3}{\alpha_3+C_{0,0,1}}}
\sqrt{1+\frac{1}{N_3}\left(\left\{\Delta^{(1)}_{3}\right\}^2\frac{\Omega^+_{123}}{\Omega_3}
+ \left\{\Delta^{(2)}_{3}\right\}^2\frac{\Omega^-_{123}}{\Omega_3}
+ \left\{\Delta^{(3)}_{3}\right\}^2\frac{\omega}{\Omega_3} - 1\right)} \times \nonumber \\
& &
\!\!\!\!\!\!
\sqrt{1+\frac{1}{N_3}\left(
\left\{\Delta^{(2)}_{1} \Delta^{(3)}_{2} - \Delta^{(2)}_{2} \Delta^{(3)}_{1}\right\}^2 \frac{\Omega_3}{\Omega^+_{123}} +
\left\{\Delta^{(1)}_{1} \Delta^{(3)}_{2} - \Delta^{(1)}_{2} \Delta^{(3)}_{1}\right\}^2 \frac{\Omega_3}{\Omega^-_{123}} +
\left\{\Delta^{(1)}_{1} \Delta^{(2)}_{2} - \Delta^{(1)}_{2} \Delta^{(2)}_{1}\right\}^2 \frac{\Omega_3}{\omega} - 1\right)}. \nonumber \
\eeqn
}}
Finally, the corresponding scalings of the natural orbitals [see in Mehler's formula above (\ref{RHO_EIGEN})] for the
different species are
$s^{(1)}_1 = \sqrt{\alpha_1(\alpha_1+C_{1,0,0})}$,
$s^{(1)}_2 = \sqrt{\alpha_2(\alpha_2+C_{0,1,0})}$,
and
$s^{(1)}_3 = \sqrt{\alpha_3(\alpha_3+C_{0,0,1})}$. 
In Sec.~\ref{EXAMPLES} we investigate the depletions of the
three species under various conditions.

\section{Illustrative examples}\label{EXAMPLES}

What governs fragmentation in a trapped multiple-species bosonic mixture?
To this end, 
in the present work we explore some answers using the Hamiltonian (\ref{HAM}).
We switch off the intraspecies interactions completely, i.e., $\lambda_1=\lambda_2=\lambda_3=0$.
In this case, in the absence of interspecies interactions the three non-interacting bosonic species are, of course, fully condensed.
In what follows, we are hence interested to look into
how interspecies interactions alone govern fragmentation,
focusing on scenarios that go beyond two-species mixtures and obviously do not exist for single-species bosons in a trap.

Two examples are put forward and investigated.
In the first, we take a two-species bosonic system (species $1$ and $3$) and embed it in a third species which serves as a bath (species $2$).
We fix the interaction within the two-species system and only alter the couplings to the bath.
Control of fragmentations of all species, the system and the bath, is demonstrated and discussed.
In the second case, we explore and exploit matters of connectivity, namely, which species interacts with which species,
and illustrate that the fragmentation of a species can be controlled by changing the interaction between the other two species.
Concretely, we consider a mixture where species $1$ interacts only with species $2$ 
which by itself also interacts with species $3$ (the interaction between species $2$ and $3$ is kept fixed).
We show that changing the way species $1$ interacts with species $2$ can drive
the fragmentation of species $3$
despite the latter not interacting directly with the former.

It is informative to examine the widths of the densities of species $1$, $2$, and $3$,
which are given by
{\footnotesize{
\beqn\label{DENS_WIDTH}
& &
\!\!\!\!\!\!
\sigma_1^{(1)}=\sqrt{\frac{1}{2\left(\alpha_1+C_{1,0,0}\right)}} = \\
& &
\!\!\!\!\!\!
\sqrt{\frac{1+\frac{1}{N_1}\left(
\left\{\Delta^{(2)}_{2} \Delta^{(3)}_{3} - \Delta^{(2)}_{3} \Delta^{(3)}_{2}\right\}^2 \frac{\Omega_1}{\Omega^+_{123}} +
\left\{\Delta^{(1)}_{2} \Delta^{(3)}_{3} - \Delta^{(1)}_{3} \Delta^{(3)}_{2}\right\}^2 \frac{\Omega_1}{\Omega^-_{123}} +
\left\{\Delta^{(1)}_{2} \Delta^{(2)}_{3} - \Delta^{(1)}_{3} \Delta^{(2)}_{2}\right\}^2 \frac{\Omega_1}{\omega} - 1\right)}{2m_1\Omega_1}}
\nonumber \
\eeqn
}}
and, analogously, by
$\sigma_2^{(1)}=\sqrt{\frac{1}{2\left(\alpha_2+C_{0,1,0}\right)}}$
and
$\sigma_3^{(1)}=\sqrt{\frac{1}{2\left(\alpha_3+C_{0,0,1}\right)}}$,
respectively.
Since all densities are Gaussian shaped, the width is simply that of a Gaussian,
$\left(\frac{1}{2\pi\sigma^2}\right)^{\frac{3}{2}}e^{-\frac{\x^2}{2\sigma^2}}$.
As done throughout this work,
the width of the many-boson density (\ref{DENS_WIDTH}) of species $1$ is an analytical closed-form
expression of all parameters in the multiple-species bosonic mixture.
Of course, this also holds for the other species.

\begin{figure}[!]
\begin{center}
\hglue -1.2 truecm
\includegraphics[width=0.36\columnwidth,angle=-90]{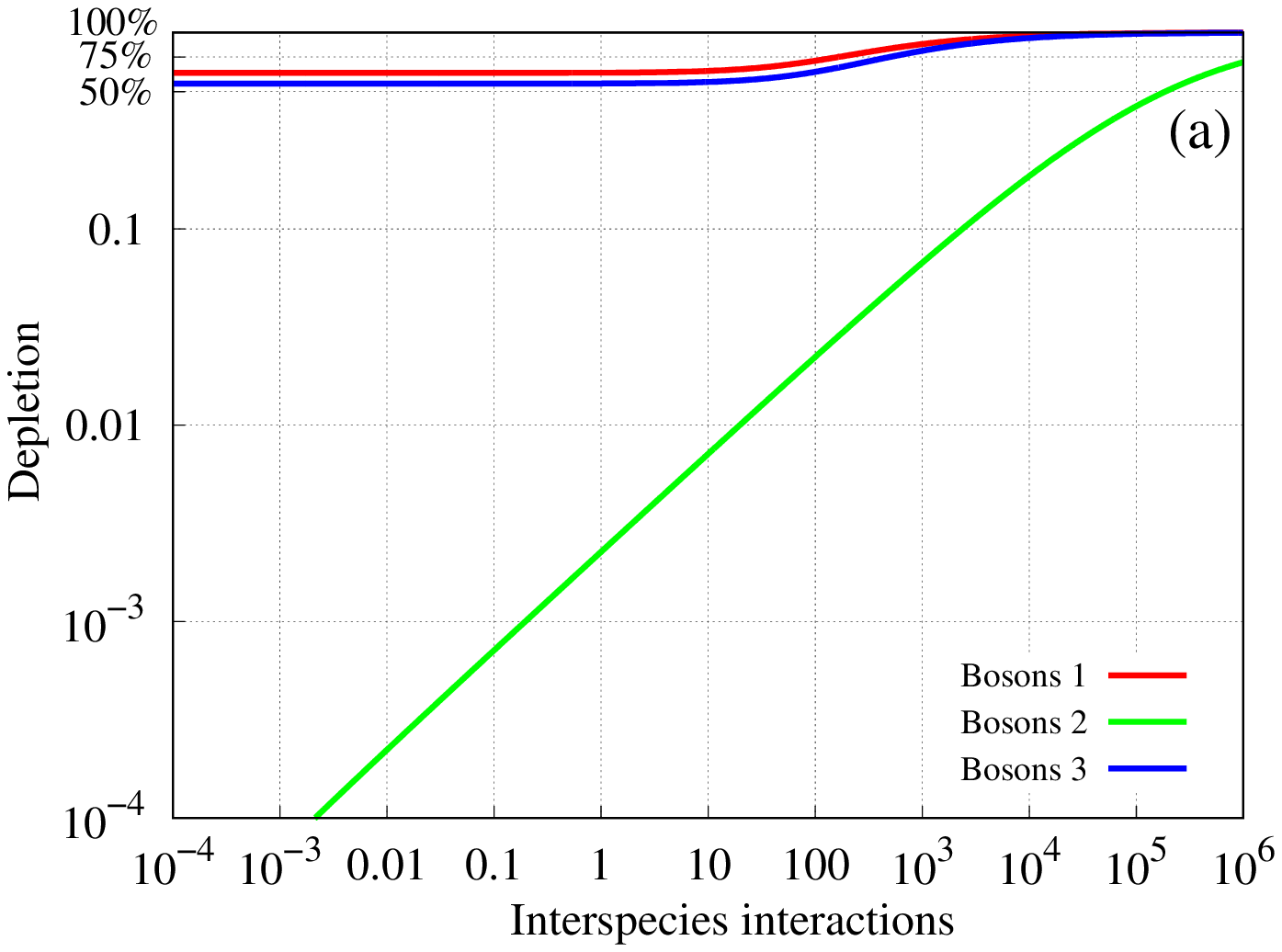}
\hglue -0.25 truecm
\includegraphics[width=0.36\columnwidth,angle=-90]{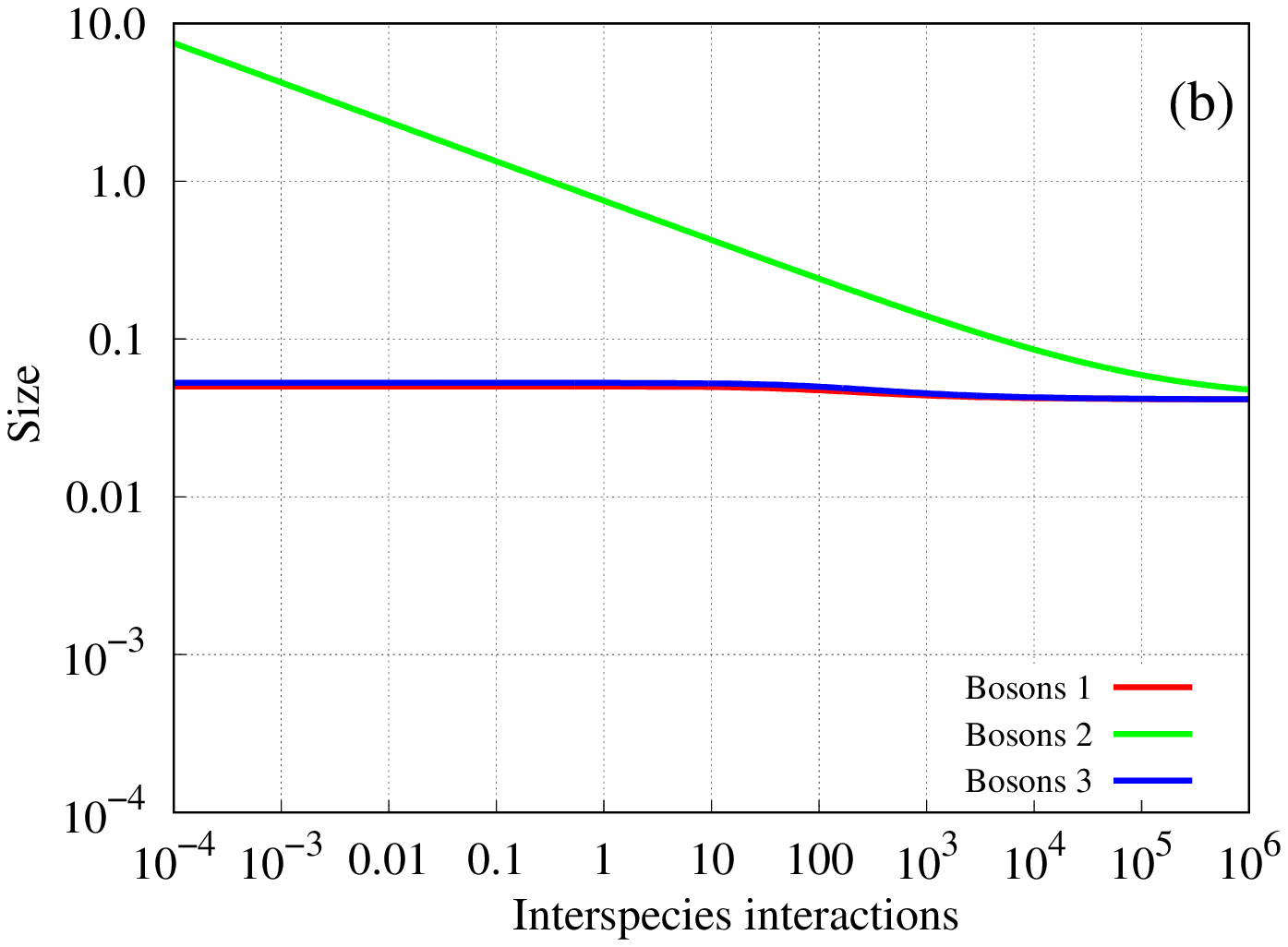}
\hglue -1.2 truecm
\includegraphics[width=0.36\columnwidth,angle=-90]{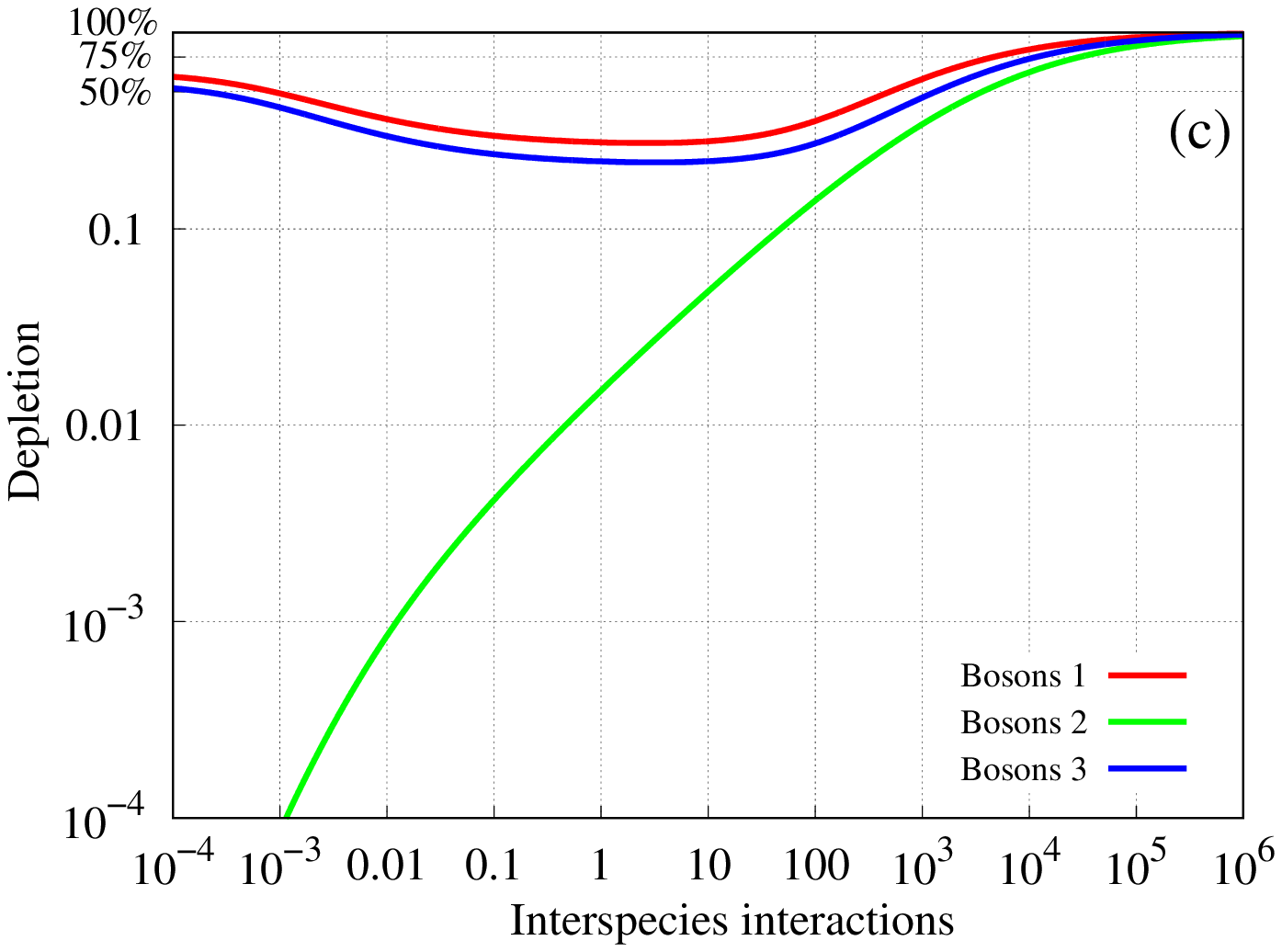}
\hglue -0.25 truecm
\includegraphics[width=0.36\columnwidth,angle=-90]{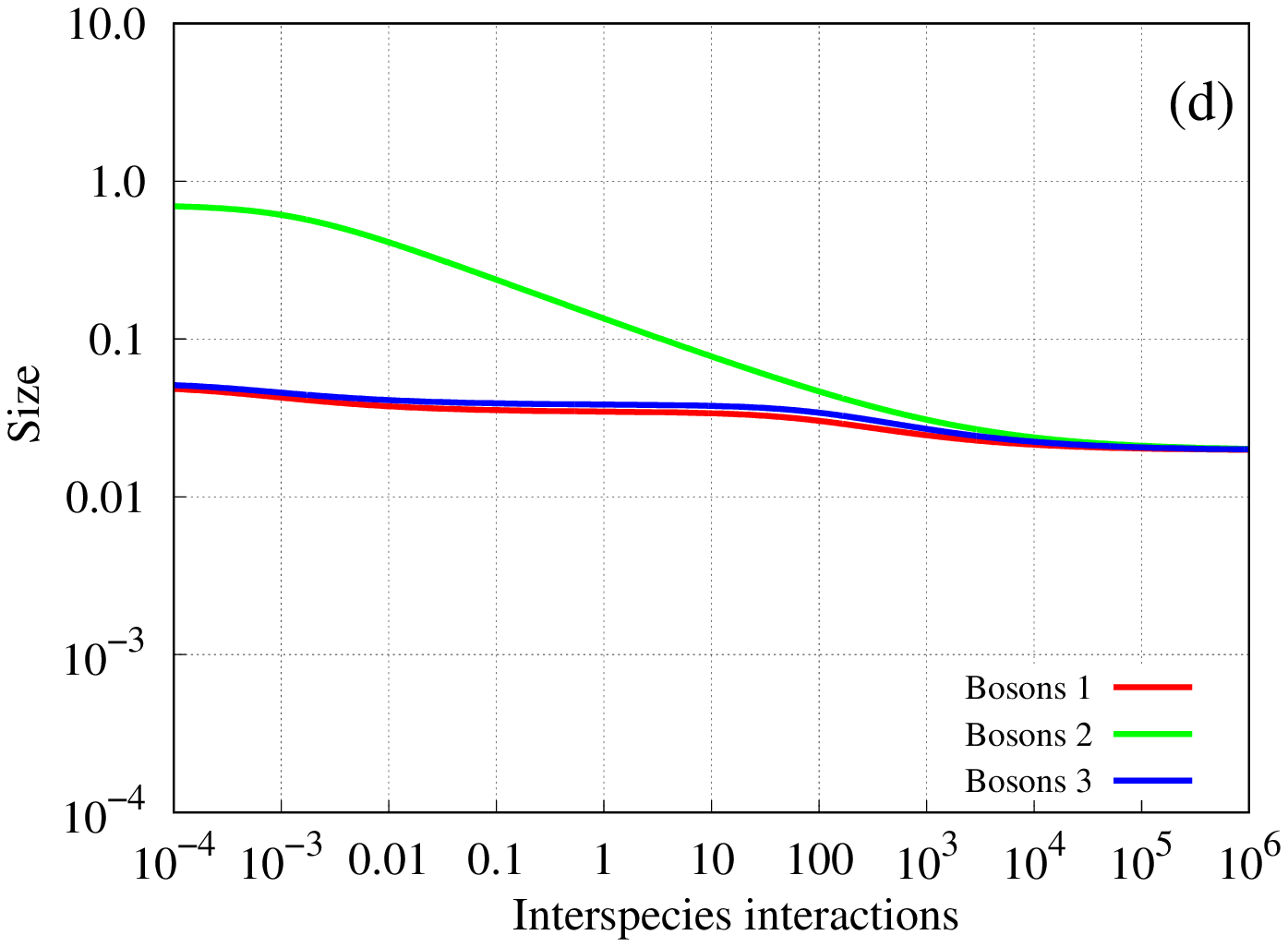}
\hglue -1.2 truecm
\includegraphics[width=0.36\columnwidth,angle=-90]{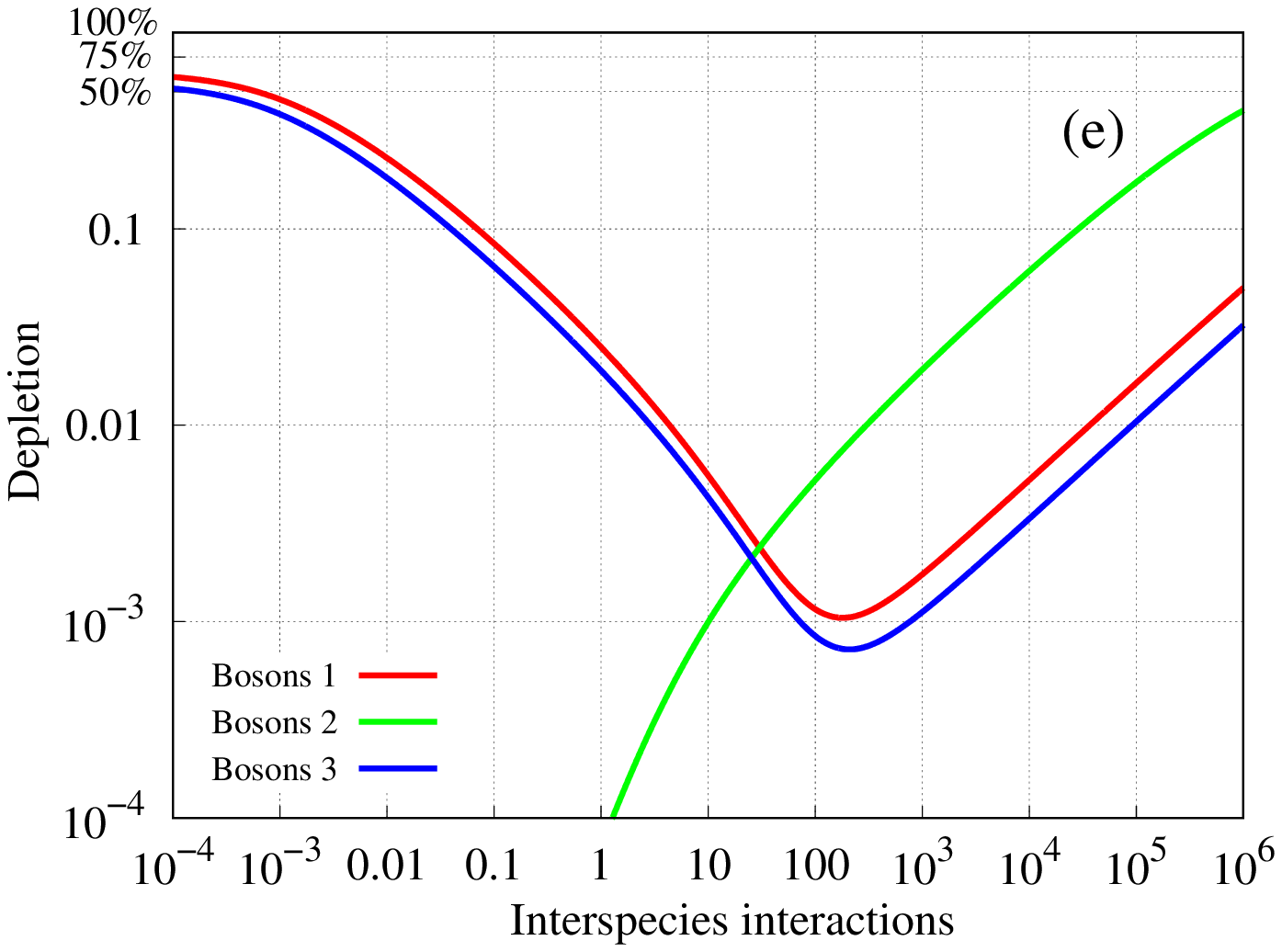}
\hglue -0.25 truecm
\includegraphics[width=0.36\columnwidth,angle=-90]{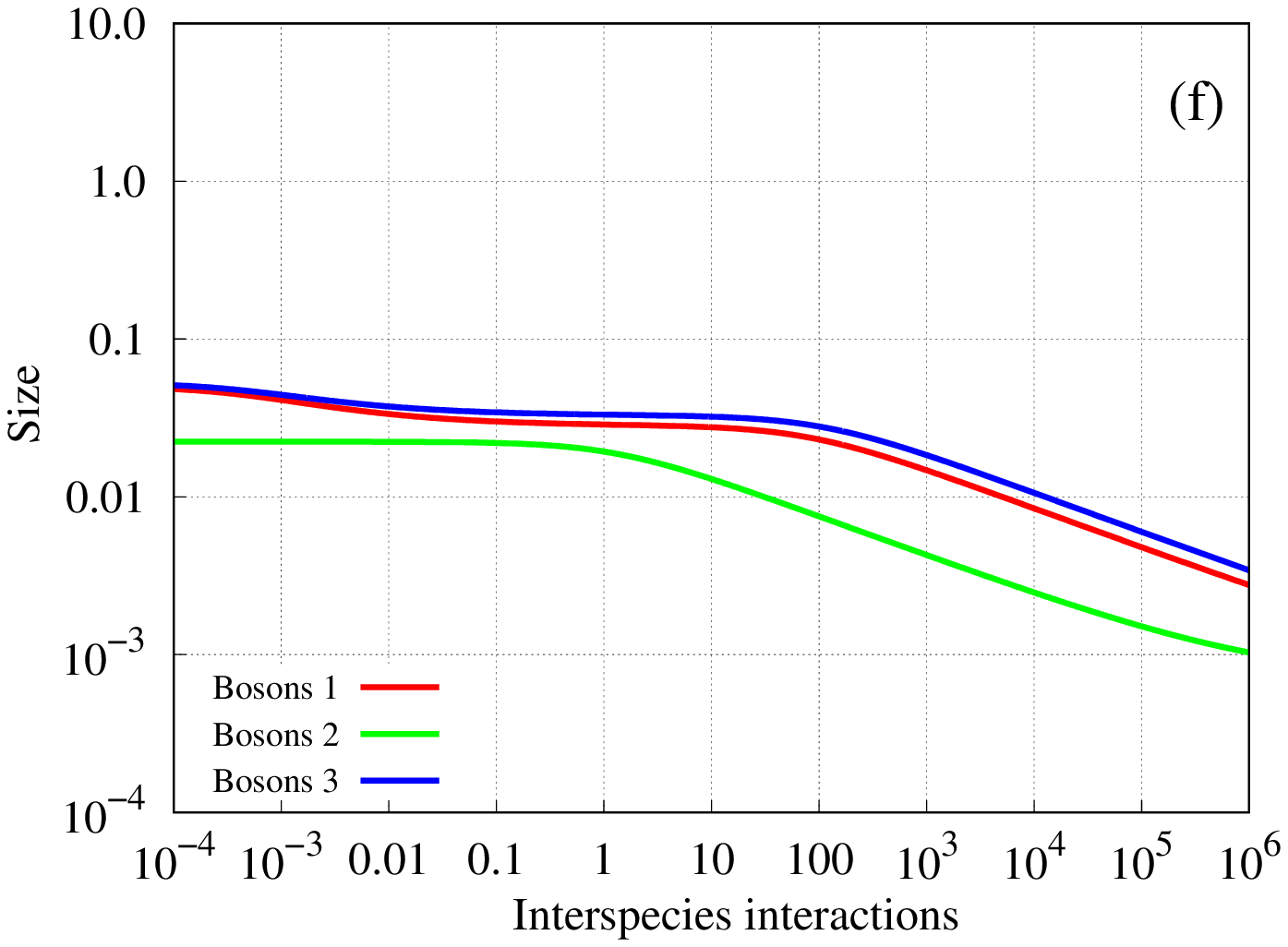}
\end{center}
\vglue 0.5 truecm
\caption{(Caption next page.)}
\label{F1}
\end{figure}

Fig.~\ref{F1} collects the results of the first investigation.
We consider a system made of $N_1=120$ bosons of type $1$,
$N_2=1000$ bosons of type $2$,
and $N_3=150$ bosons of type $3$.
In this example species $2$ acts
as a bath for the system which is made of species $1$ and $3$.
The respective masses are $m_1=1.3$, $m_2=1.0$, and $m_3=0.9$.
We also present a comparison for different masses of the bath bosons,
namely, $m_2=0.001$ and $m_2=1000.0$,
which is found to be appealing, see below.
Respective parameters are chosen to differ from each other,
to emphasize the
treatment of a generic, imbalanced mixture.

The corresponding interspecies interactions are $\lambda_{12}=1.9\lambda$,
$\lambda_{13}=1000.0$,
and $\lambda_{23}=1.1\lambda$.
In other words,
we keep the interaction between species $1$ and $3$, the system, constant
while jointly and proportionally
increasing the couplings of species $1$ and $3$ to the bath, species $2$,
by varying $\lambda$.
The intraspecies interactions, as said above, vanish in the following examples.

In the absence of couplings to the bath, the fragmentations
of species $1$ and $3$ are, respectively,
$d^{(1)}_1 = 0.6234$ and
$d^{(1)}_3 = 0.5494$ (no rounding here and hereafter).
Correspondingly, the sizes of their densities are
$\sigma_1^{(1)}=0.05014$ and
$\sigma_3^{(1)}=0.05281$.
Here and hereafter, we use size and width of a density interchangeably.
Overall, see Fig.~\ref{F1},
an intriguing dependence of the depletions on interactions with the bath
as well as on the mass of the bosons comprising the bath is found.
Generally, 
increasing the couplings between the two-species system and the bath
increases the fragmentation of the bath monotonously.
Yet, the fragmentations of species $1$ and $3$ behave in a non-monotonous manner.
Here,
the larger is the mass of the bath, the stronger is the variation of the system's fragmentation
which, first, decreases more and more and only for stronger couplings increases.
Interestingly,
the depletion of the bath is non-monotonous with the mass of the bath,
i.e., given the interactions between the system and the bath,
there is an optimal mass for the maximal depletion of the bath.
Such a behavior of the fragmentation of the bath
is not found in the corresponding two-species mixture \cite{BB_2023}.

An accompanying analysis of the respective sizes of the three bosonic clouds is instrumental.
Here, generally, increasing the couplings
to the bath decreases its size as well as that of the system's species,
see Fig.~\ref{F1}.
Yet, when the mass of the bath is larger,
the sizes of species $1$ and $3$ are more affected by species $2$ upon enlarging the interactions.
Correspondingly, when the size of the bath is much larger than that of the system,
enlarging the couplings leads essentially to only increase in the fragmentations of the system.
When the mass of the bath is increased, the size of the bath obviously decreases,
and the effect on the system is to first lower its
fragmentations and later on to increase the fragmentations,
see also the discussion above.
All together, the changes in the sizes of the system and the bath is much more involved in the three-species mixture
than in the two-species mixture \cite{BB_2023}.
We emphasize that all results are expressed analytically explicitly
in closed form.

\addtocounter{figure}{-1}
\begin{figure}[!]
\begin{center}
\caption{(Previous page.)
Fragmentation of a three-species bosonic mixture: Two-species system coupled to a third-species bath.
The left column depicts the 
depletions
of bosons in a mixture made of
$N_1=120$ species $1$ and $N_3=150$ species $3$ bosons
(the system) coupled to $N_2=1000$ species $2$ bosons (the bath)
as a function of the interspecies interaction strengths.
Here, $\lambda_{13}=1000.0$ is kept fixed,
and $\lambda_{12}=1.9\lambda$ and $\lambda_{23}=1.1\lambda$ are increased linearly
($\lambda$ is the common x-axis in the panels).
The intraspecies interactions of all species are zero.
The sizes of each bosonic cloud are depicted in the right column.
The masses are
$m_1=1.3$, $m_3=0.9$,
and (a,b) $m_2=0.001$; (c,d) $m_2=1.0$; and (e,f) $m_2=1000.0$.
An intriguing dependence of the depletions and sizes on the interactions and masses is found.
All results are obtained analytically.
See the text for more details.
The quantities shown are dimensionless.
}
\end{center}
\end{figure}

\begin{figure}[!]
\begin{center}
\hglue -1.2 truecm
\includegraphics[width=0.36\columnwidth,angle=-90]{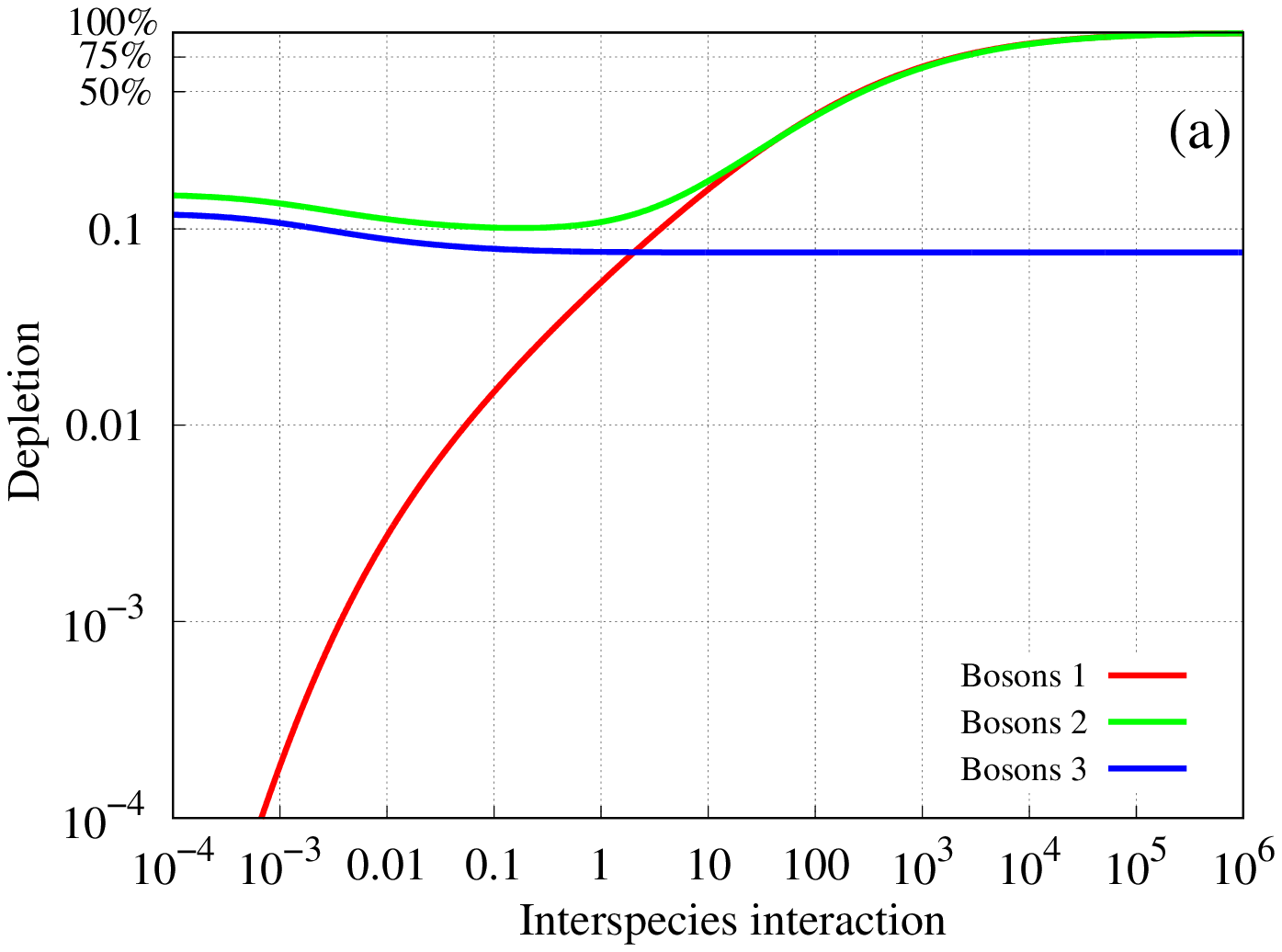}
\hglue -0.25 truecm
\includegraphics[width=0.36\columnwidth,angle=-90]{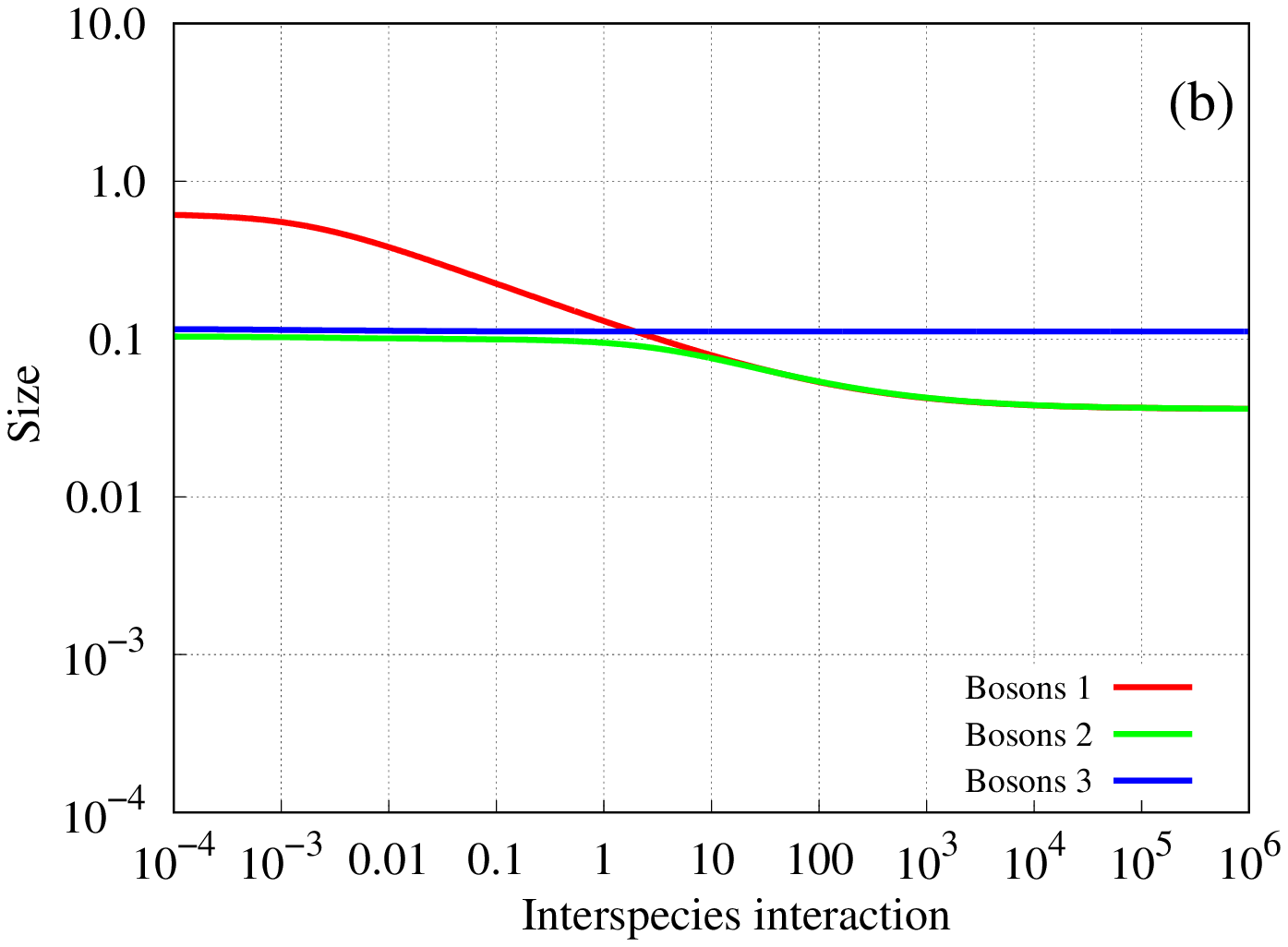}
\hglue -1.2 truecm
\includegraphics[width=0.36\columnwidth,angle=-90]{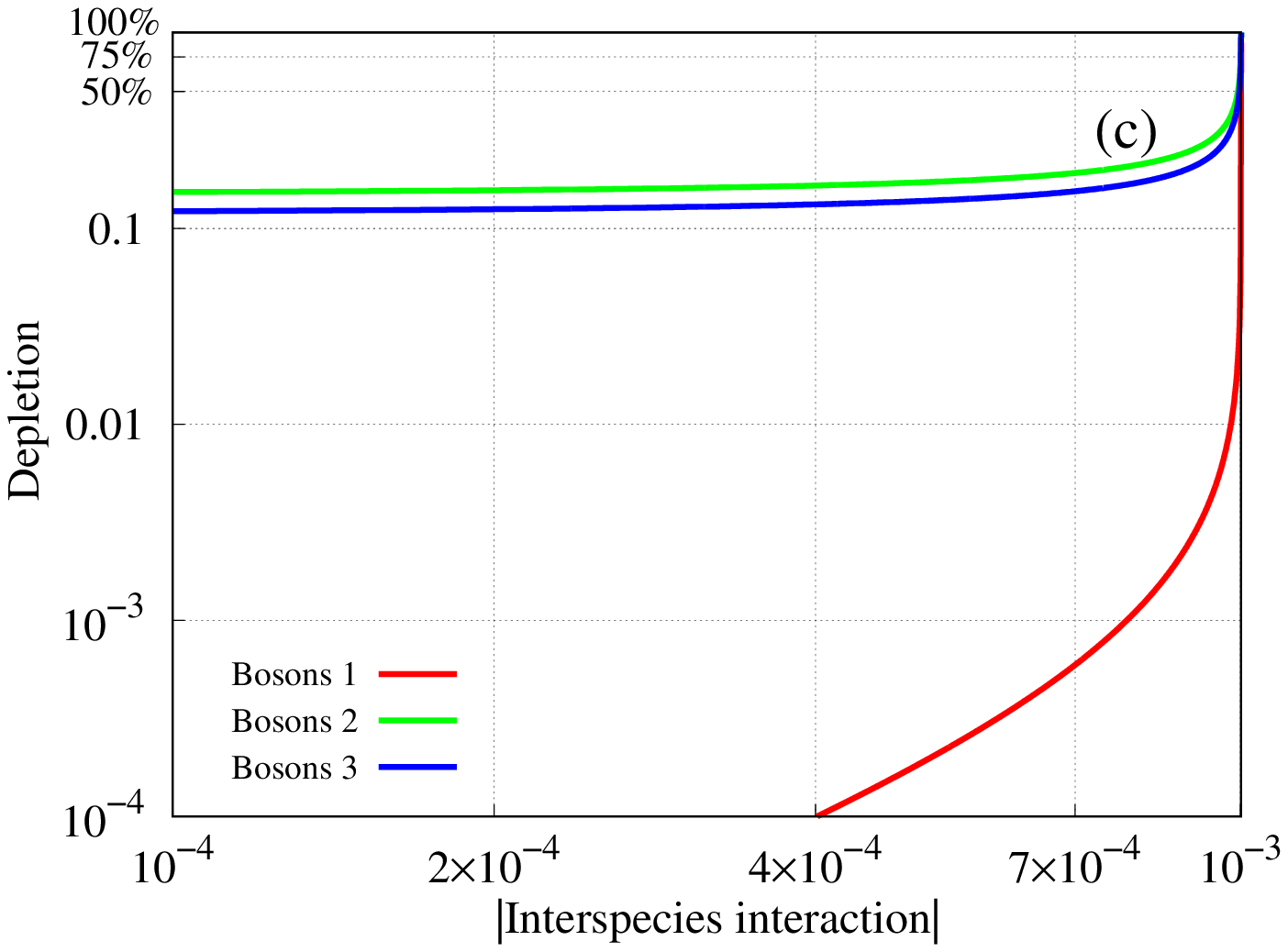}
\hglue -0.25 truecm
\includegraphics[width=0.36\columnwidth,angle=-90]{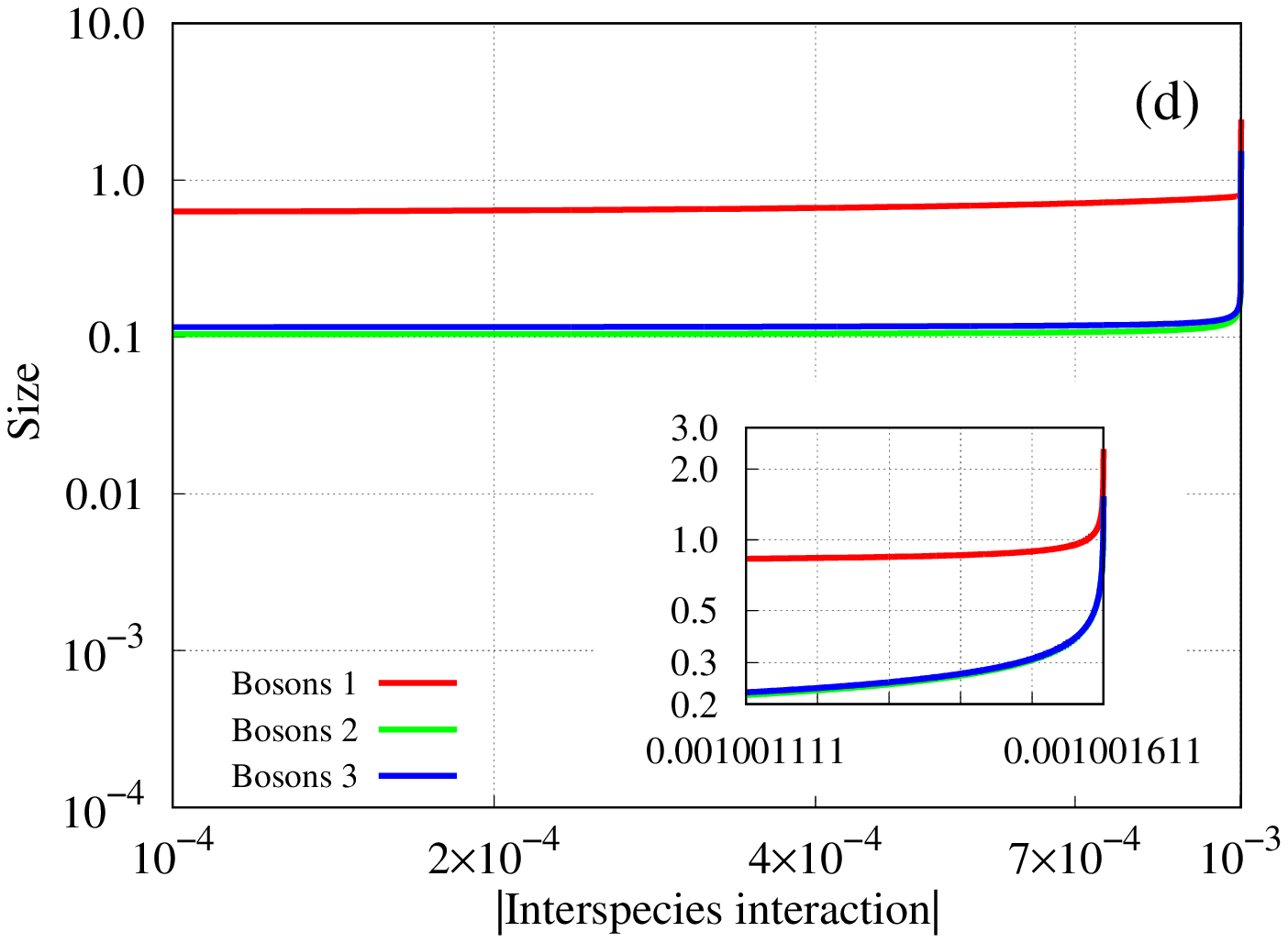}
\end{center}
\vglue 0.5 truecm
\caption{Fragmentation of a three-species bosonic mixture: Effect of connectivity.
The left column depicts the 
depletions
of bosons in a mixture made of
$N_1=120$ species $1$, $N_2=100$ species $2$, and $N_3=150$ species $3$ bosons
as a function of an interspecies interaction strength.
Here, $\lambda_{12}=3.9\lambda$ is varied
linearly ($\lambda$ is the x-axis in the panels),
$\lambda_{23}=11.0$ is held fixed,
and
there is no interaction between species $1$ and $3$,
$\lambda_{13}=0.0$.
The intraspecies interactions of all species are zero.
The sizes of each bosonic cloud are depicted in the right column.
The masses are
$m_1=1.3$, $m_2=1.0$, and $m_3=0.9$.
An intricate dependence of the fragmentations and sizes on the interaction is identified.
All results are obtained analytically.
See the text for more details.
The quantities shown are dimensionless.}
\label{F2}
\end{figure}

Fig.~\ref{F2} depicts the results of the second investigation.
Now, the numbers of particles are 
$N_1=120$, $N_2=100$, and $N_3=150$
and their masses
$m_1=1.3$, $m_2=1.0$, and $m_3=0.9$, respectively.
We fix the interaction between species $2$ and $3$, $\lambda_{23}=11.0$,
while the interaction between species $1$ and $3$ vanishes, $\lambda_{13}=0.0$.
Only the interaction strength between species $1$ and $2$ is varied,
$\lambda_{12}=3.9\lambda$.
As in the previous example,
the intraspecies interactions of all species are zero.

In the absence of interaction between species $1$ and $2$,
the depletions
of species $2$ and $3$ are, correspondingly, 
$d^{(1)}_2 = 0.1504$ and
$d^{(1)}_3 = 0.1201$.
The respective sizes are
$\sigma_2^{(1)}=0.1038$ and
$\sigma_3^{(1)}=0.1153$.
The fragmentations and sizes of each bosonic cloud 
are computed in the present example
both for attractive and repulsive interspecies interaction $\lambda_{12}$,
when the mixture is still bound, see below.
Since the sign of $\lambda_{12}$ is positive on the attractive branch and negative on the repulsive branch,
to avoid confusion,
referring to increasing the interaction implies increasing the magnitude of the interaction.
An intriguing dependence of the depletions and sizes on the interaction is found, see Fig.~\ref{F2}.
Again, all results are obtained analytically and expressed
explicitly using the theory derived above.

On the attractive branch, see upper panels of Fig.~\ref{F2},
increasing the interaction between species $1$ and $2$
eventually leads to a common growth of their fragmentations and joint decrease and subsequent
tendency for saturation of their sizes.
Species $3$, which `grows apart' from species $2$ when the interaction is enlarged,
exhibits a small lowering of its depletion and a very mild decrease in its size,
both tending saturation.

On the repulsive branch, the picture differs qualitatively.
Enlarging the repulsion between species $1$ and $2$ leads to a steady increase
in the fragmentations of all species and growth of their sizes,
see lower panels of Fig.~\ref{F2}.
In particular,
approaching the border of stability the rise of depletions and sizes
of all species `accelerates' and tends to their limiting values of unity and infinity,
respectively. 
Thus, species $3$ becomes more and more fragmented because
species $1$ `pushes' species $2$ which is coupled to species $3$.

Consider the interaction strength
$\lambda_{12}= - 3.9 \times 0.001001611$,
where the mixture is just about to become unbound.
Here, the fragmentations are already
$d^{(1)}_1=0.8850$,
$d^{(1)}_2=0.9984$,
and
$d^{(1)}_3=0.9978$.
The corresponding widths are
$\sigma_1^{(1)}=2.432$,
$\sigma_2^{(1)}=1.533(6)$,
and
$\sigma_3^{(1)}=1.533(8)$,
compare to their respective values for $\lambda_{12}=0.0$
[$\sigma_1^{(1)}=0.6201$, $\sigma_2^{(1)}=0.1038$, and $\sigma_3^{(1)}=0.1153$].
How close are we to the border of stability, i.e.,
when at least one of the frequencies (\ref{REL_OMG}) and (\ref{CM_OMG}) becomes zero, also see (\ref{FREQ_COND}),
for $\lambda_{12}= - 3.9 \times 0.00100161{\underline 1}$?
The intraspecies relative-motion frequencies are found to be $\Omega_1=0.6316$, $\Omega_2=57.44$, and $49.45$,
i.e., they are not the reason for approaching the border of stability.
On the other end, examining the relative-motion center-of-mass frequencies it is found that,
whereas $\Omega^+_{123}=75.79$,
we have $\Omega^-_{123}=0.0003625$.
Hence, for the example studied here approaching the border of stability is
governed by softening of a relative-motion center-of-mass coordinate.
Finally, we note
that for $\lambda_{12}= - 3.9 \times 0.00100161{\underline 2}$ the mixture is already unbound.
All in all, altering the interaction between species $1$ and $2$
impacts crucially the properties of species $3$,
despite having fixed interaction with species $2$
and no interaction with species $1$.
This is a proper place to conclude the present investigation.

\section{Summary and outlook}\label{SUMMARY}

In the current
work we investigated fragmentation of a multiple-species
bosonic mixture from the point of an exactly-solvable many-particle model.
Such models for multiple-species trapped bosons are rather scarce,
making the solution presented here of particular aesthetics and interest on its own.

We solved the many-particle Hamiltonian and investigated the structure of the
ground-state wavefunction and energy and their dependences on all parameters,
the masses, numbers of particles, and the intraspecies and interspecies interaction strengths.
Tools to integrate the reduced one-particle density matrices for all the species
were put forward,
and the explicit expressions as a function of all parameters in the mixture were prescribed
and analyzed.
As illustrative examples,
we focused on scenarios that go beyond the respective single-species and two-species systems,
and highlighted intricate fragmentation properties of the species
that only occur in the multiple-species mixture.

As an outlook, we touch upon the flexibilities and opportunities offered by the generic multiple-species
harmonic-interaction model.
These include systems with impurities, induced interactions between mutually non-interacting species coupled to another species and,
further down the road,
out-of-equilibrium dynamics, all from the point of view of analytical many-body results.
We also believe that more insights into the connections and differences between the
many-body and mean-field solutions are achievable beyond \cite{JCP_2024},
i.e., benefiting from the imbalance between species.
Methodologically, further tools to handle various reduced-density matrices would be instrumental.
Another facet that can be anticipated,
is using the plethora of analytical results to assess the regimes of validity of numerical approaches.
Also, extensions of the model for different harmonic trappings of each species and distinct positions of their minima might prove useful.
Finally, applications to bosons in cavities, whether multiple-species bosons in a single-mode cavity or the other way around,
in the spirit of \cite{HIM_BEC_CAVITY},
could be foreseen.

\section*{Acknowledgements}

This research was supported by the Israel Science Foundation (Grant No. 1516/19). 

\appendix

\section{Limiting and specific cases of the frequencies' matrix and its eigenvectors and eigenvalues}\label{APP_LIMITS}

In what follows we denote the eigenvectors (\ref{BBB_CM_REL}) and (\ref{BBB_CM_ALL}) of the frequencies' matrix as 
${\bf \Delta}^{(J)}=\left(\Delta^{(J)}_{1},\Delta^{(J)}_{2},\Delta^{(J)}_{3}\right)\!, \, J=1,2,3$ 
and enlist and discuss several possibilities.

\begin{enumerate}
\item[A1.] Generally, the three interspecies interactions $\lambda_{12}$, $\lambda_{13}$, and $\lambda_{23}$  
are nonzero and couple the three species.
We have then a generic three-species mixture where
the relative-motion center-of-mass coordinates (\ref{BBB_CM_REL})
depend on the interspecies interactions and the frequencies $\{\Omega_{123}^+,\Omega_{123}^-,\omega\}$ are non-degenerate.
Scarcely, the explicit expression for a relative-motion center-of-mass coordinate in (\ref{BBB_CM_REL}) may have
an accidental vanishing point;
for instance, at $m_1=2m$, $m_2=m_3=m$ and $\lambda_{13}=2\lambda$, $\lambda_{12}=\lambda_{23}=\lambda$, $\lambda>0$,
we have ${\bf \Delta}^{(2)} \equiv 0$ for any numbers $N_1$, $N_2$, and $N_3$ of bosons.
In such a case, 
its value at the accidental vanishing point is determined as a limit, specifically, say, $\lambda_{23} \to \lambda^+$.
The result, which recovers of course that obtained directly by diagonalizing the frequencies' matrix at this point,
is ${\bf \bar \Delta}^{(2)} = \frac{1}{\sqrt{2N_1+N2}}\left(\sqrt{N_2},-\sqrt{2N_1},0\right)$.
Interestingly, it does not depend on $N_3$,
the number of bosons of species $3$.

\item[A2.] If one interspecies interaction is zero,
we shall choose the interaction between species $1$ and species $3$ to vanish,
i.e., $\lambda_{13}=0$.
Correspondingly, the above expressions for $K$ and $G$ (\ref{CM_3_freq_KG}) simplify a bit.
Indeed, substituting $\Lambda_{13},\Lambda_{31}=0$ into the three-body part $G$
the latter does not vanish, namely, we still have a three-species mixture,
the relative-motion center-of-mass coordinates depend on the interspecies interactions,
and the frequencies $\{\Omega_{123}^+,\Omega_{123}^-,\omega\}$ are non-degenerate.

\item[A3.] If two interspecies interaction are zero,
say $\lambda_{12}=0$ and $\lambda_{23}=0$,
then the three-body part vanishes, $G=0$, and consequently
$\Omega_{123}^-$
degenerates
with the trap frequency $\omega$.
Physically, the system boils down to a mixture of two species, $1$ and $3$, and one individual species, $2$,
and therefore the expressions for the two-species mixtures \cite{HIM_MIX_RDM} plus one individual species should be used.
Explicitly, we have for the three species
${\bf \bar \Delta}^{(1)}=\left(-\sqrt{\frac{m_3N_3}{m_1N_1+m_3N_3}},0,\sqrt{\frac{m_1N_1}{m_1N_1+m_3N_3}}\right)$,
${\bf \bar \Delta}^{(2)}=\left(0,1,0\right)$,
and
${\bf \bar \Delta}^{(3)}=\left(\sqrt{\frac{m_1N_1}{m_1N_1+m_3N_3}},0,\sqrt{\frac{m_3N_3}{m_1N_1+m_3N_3}}\right)$.
Indeed, taking $\lambda_{12}=\lambda_{23}=\lambda$, with $\lambda \to 0^+$,
and for $\lambda_{13}>0$,
Eq.~(\ref{BBB_CM_REL}) becomes
${\bf \Delta}^{(1)}={\bf \bar \Delta}^{(1)}$
and 
${\bf \Delta}^{(2)}=\sqrt{\frac{m_2N_2}{(m_1N_1+m_3N_3)((m_1N_1+m_2N_2+m_3N_3)}}\break\hfill
\left(-\sqrt{m_1N_1},\frac{m_1N_1+m_3N_3}{\sqrt{m_2N_2}},-\sqrt{m_3N_3}\right)$, respectively,
and
${\bf \Delta}^{(3)}$
is simply taken from (\ref{BBB_CM_ALL}).
Thus,
the (remaining) relative-motion center-of-mass coordinate ${\bf \Delta}^{(1)}$ becomes\break\hfill interaction-independent now. 
Then,
taking a linear combination of eigenvectors associates with the degenerate eigenvalue $\omega$,
${\bf \Delta}^{(2)}$ and ${\bf \Delta}^{(3)}$,
gives the remaining physical center-of-mass coordinates
${\bf \bar \Delta}^{(2)}$ and ${\bf \bar \Delta}^{(3)}$.
All in all, the frequencies are
$\{\Omega_{123}^+=\sqrt{\omega^2+2\left(\frac{\Lambda_{13}}{m_3}+\frac{\Lambda_{31}}{m_1}\right)},\Omega_{123}^-=\omega,\omega\}$.

\item[A4.] If all three interspecies interactions are zero
then $\Omega_{123}^\pm=\omega$.
Physically,
one deals of course with three individual species
where the frequencies' matrix (\ref{HAM_MIX_3_XYZ_CMs})
is diagonal
and its eigenvectors
trivial,
${\bf \bar \Delta}^{(1)}=\left(1,0,0\right)$,
${\bf \bar \Delta}^{(2)}=\left(0,1,0\right)$,
and ${\bf \bar \Delta}^{(3)}=\left(0,0,1\right)$,
i.e.,
all coordinates become species' center-of-mass coordinates.
To recover the limit of three individual species from the generic solution,
the previous scenario (A3.) is performed first, i.e., we start with a mixture of $1$ and $3$ for which
${\bf \Delta}^{(1)}=\left(-\sqrt{\frac{m_3N_3}{m_1N_1+m_3N_3}},0,\sqrt{\frac{m_1N_1}{m_1N_1+m_3N_3}}\right)$
and
${\bf \Delta}^{(3)}=\left(\sqrt{\frac{m_1N_1}{m_1N_1+m_3N_3}},0,\sqrt{\frac{m_3N_3}{m_1N_1+m_3N_3}}\right)$,
and the individual species $2$.
Then, taking $\lambda_{13} \to 0^+$ means
$\Omega_{123}^+$ degenerates with the trap frequency $\omega$ as well,
implying that one can mix 
${\bf \Delta}^{(1)}$ and ${\bf \Delta}^{(3)}$
to get the
physical center-of-mass coordinates ${\bf \bar \Delta}^{(1)}$ and ${\bf \bar \Delta}^{(3)}$.
In sum,
the
limit of a non-interacting mixture whose
frequencies are three-fold degenerate, $\{\Omega_{123}^+=\omega,\Omega_{123}^-=\omega,\omega\}$,
is readily obtained from
the generic three-species expressions
(\ref{BBB_CM_REL}) and (\ref{BBB_CM_ALL}).

\item[A5.] The two relative-motion center-of-mass roots
become degenerate in the specific case 
of a mass-balanced and interaction-balanced three-species mixture.
The numbers of bosons $N_1$, $N_2$, and $N_3$ need not be equal,
but when they are,
we have the very specific case of a balanced three-species mixture dealt with in \cite{JCP_2024}.
Denoting by $m$ all masses and $\lambda$ all interspecies interaction strengths,
we have
$\{\Omega_{123}^+=\sqrt{\omega^2 + \frac{2\lambda}{m}\left(N_1+N_2+N_3\right)},\Omega_{123}^-=\sqrt{\omega^2 + \frac{2\lambda}{m}\left(N_1+N_2+N_3\right)},\omega\}$.
The respective
eigenvectors do not depend on the interactions and are
${\bf \bar \Delta}^{(1)}=\frac{1}{\sqrt{(N_1+N_3)(N_1+N_2+N_3)}}\break\hfill
\left(\sqrt{N_1N_2},-(N_1+N_3),\sqrt{N_2N_3}\right)$ and
${\bf \bar \Delta}^{(2)}=\left(\sqrt{\frac{N_3}{N_1+N_3}},0,-\sqrt{\frac{N_1}{N_1+N_3}}\right)$
along with the center-of-mass coordinate
${\bf \bar \Delta}^{(3)}=\frac{1}{\sqrt{N_1+N_2+N_3}}
\left(\sqrt{N_1},\sqrt{N_2},\sqrt{N_3}\right)$.
Now, substituting in (\ref{BBB_CM_REL}) and (\ref{BBB_CM_ALL}) $m$ for all masses, $\lambda_{12}=\lambda_{23}=\lambda$, for $\lambda>0$,
and taking the limit $\lambda_{13} \to \lambda^{-}$,
one recovers ${\bf \Delta}^{(1)}={\bf \bar \Delta}^{(1)}$,
${\bf \Delta}^{(2)}={\bf \bar \Delta}^{(2)}$,
and
${\bf \Delta}^{(3)}={\bf \bar \Delta}^{(3)}$
for the eigenvectors of the center-of-masses Hamiltonian.
Note that the degenerate frequencies $\Omega_{123}^\pm$ depend on the sum of all bosons only
and one might suspect that the mixture behaves as a single-species mixture.
This is not the case.
As can be seen,
the fragmentations of the species do depend on the individual numbers of bosons $N_1$, $N_2$, $N_3$.
\end{enumerate}

\section{The mean-field solution of the generic multiple-species mixture}\label{MF}

The mean-field solution of the three-species mixture goes as follows.
The generalization to any number of species $P$ is straightforward.
The ansatz for the wavefunction is the separable product state
\beq\label{MIX_WAV_GP_3}
 \Phi^{GP}(\x_1,\ldots,\x_{N_1},\y_1,\ldots,\y_{N_2},\z_1,\ldots,\z_{N_3}) = 
\prod_{j=1}^{N_1} \phi^{GP}_1(\x_j) \prod_{k=1}^{N_2} \phi^{GP}_2(\y_k) \prod_{l=1}^{N_3} \phi^{GP}_3(\z_l), 
\eeq
where all bosons of the same type reside in one and the same normalized orbital.
Sandwiching the Hamiltonian (\ref{HAM})
with the ansatz (\ref{MIX_WAV_GP_3})
and
minimizing the resulting energy functional with respect to
the shapes of the orbitals $\phi_1(\x)$, $\phi_2(\y)$, and $\phi_3(\z)$, 
the three-coupled Gross-Pitaevskii equations of the mixture are derived,
\beqn\label{MIX_EQ_GP_BBB_3}
& &
\bigg\{-\frac{1}{2m_1} \frac{\partial^2}{\partial \x^2} + \frac{1}{2} m_1 \omega^2 \x^2 
+ \int d\x' \big[\Lambda_1 |\phi_1(\x')|^2 + \Lambda_{21} |\phi_2(\x')|^2 + \nonumber \\
& & + \Lambda_{31} |\phi_3(\x')|^2\big] (\x-\x')^2\bigg\} \phi_1(\x) = \mu_1 \phi_1(\x),
\nonumber \\
& &
\bigg\{-\frac{1}{2m_2} \frac{\partial^2}{\partial \y^2} + \frac{1}{2} m_2 \omega^2 \y^2 
+ \int d\y' \big[\Lambda_2 |\phi_2(\y')|^2 + \Lambda_{12} |\phi_1(\y')|^2 + \nonumber \\
& & + \Lambda_{32} |\phi_3(\y')|^2\big] (\y-\y')^2\bigg\} \phi_2(\y) = \mu_2 \phi_2(\y),
\nonumber \\
& &
\bigg\{-\frac{1}{2m_3} \frac{\partial^2}{\partial \z^2} + \frac{1}{2} m_3 \omega^2 \z^2 
+ \int d\z' \big[\Lambda_3 |\phi_3(\z')|^2 + \Lambda_{13} |\phi_1(\y')|^2 + \nonumber \\
& & + \Lambda_{23} |\phi_3(\z')|^2\big] (\z-\z')^2\bigg\} \phi_3(\z) = \mu_3 \phi_3(\z), \
\eeqn
where $\mu_1$, $\mu_2$, and $\mu_3$ are the chemical potentials of the species.
The solution to the three coupled non-linear equations (\ref{MIX_EQ_GP_BBB_3}) is given by
\beqn\label{MIX_GP_OR_3}
& & \phi^{GP}_1(\x) = \left(\frac{m_1\Omega_1^{GP}}{\pi}\right)^{\frac{3}{4}}
e^{-\frac{1}{2}m_1\Omega_1^{GP}\x^2}, \quad
\Omega_1^{GP} = \sqrt{\omega^2 + \frac{2}{m_1}(\Lambda_1+\Lambda_{21}+\Lambda_{31})}, \nonumber \\
& & \phi^{GP}_2(\y) = \left(\frac{m_2\Omega_2^{GP}}{\pi}\right)^{\frac{3}{4}}
e^{-\frac{1}{2}m_2\Omega_2^{GP}\y^2}, \quad
\Omega_2^{GP} = \sqrt{\omega^2 + \frac{2}{m_2}(\Lambda_2+\Lambda_{12}+\Lambda_{32})}, \nonumber \\
& & \phi^{GP}_3(\z) = \left(\frac{m_3\Omega_1^{GP}}{\pi}\right)^{\frac{3}{4}}
e^{-\frac{1}{2}m_3\Omega_3^{GP}\z^2}, \quad
\Omega_3^{GP} = \sqrt{\omega^2 + \frac{2}{m_3}(\Lambda_3+\Lambda_{13}+\Lambda_{23})}. \
\eeqn
We find that the density of each species,
$N_1\left|\phi^{GP}_1(\x)\right|^2$,
$N_2\left|\phi^{GP}_2(\y)\right|^2$,
and
$N_3\left|\phi^{GP}_3(\z)\right|^2$,
is dressed by the interspecies interactions with the other two species,
but does not depend on the mutual
interaction between the other two species.
For completeness,
the chemical potentials read
\beqn\label{MUS}
& &
\mu_1 = \frac{1}{2}\left(\Omega_1^{GP} + \frac{\Lambda_1}{\Omega_1^{GP}} + \frac{\Lambda_{21}}{\Omega_2^{GP}} +
\frac{\Lambda_{31}}{\Omega_3^{GP}}\right), \quad 
\mu_2 = \frac{1}{2}\left(\Omega_2^{GP} + \frac{\Lambda_2}{\Omega_2^{GP}} + \frac{\Lambda_{12}}{\Omega_1^{GP}} +
\frac{\Lambda_{32}}{\Omega_3^{GP}}\right), \nonumber \\
& &
\mu_3 = \frac{1}{2}\left(\Omega_3^{GP} + \frac{\Lambda_3}{\Omega_3^{GP}} + \frac{\Lambda_{13}}{\Omega_1^{GP}} +
\frac{\Lambda_{23}}{\Omega_2^{GP}}\right). \
\eeqn

Finally, the Gross-Pitaevskii energy per particle
of the three-species mixture reads
\beqn\label{MIX_GP_E_N_3}
& &
\varepsilon^{GP} = \frac{3}{2}\frac{N_1\Omega_1^{GP} + N_2\Omega_2^{GP} + N_3\Omega_3^{GP}}{N} =
\frac{3}{2}\left[\frac{\sqrt{\omega^2 + \frac{2}{m_1}(\Lambda_1+\Lambda_{21}+\Lambda_{31})}}
{1+\frac{\Lambda_{21}}{\Lambda_{12}}+\frac{\Lambda_{31}}{\Lambda_{13}}}\right. + \nonumber \\
& & + \left.\frac{\sqrt{\omega^2 + \frac{2}{m_2}(\Lambda_2+\Lambda_{12}+\Lambda_{32})}}
{1+\frac{\Lambda_{12}}{\Lambda_{21}}+\frac{\Lambda_{32}}{\Lambda_{23}}} +
\frac{\sqrt{\omega^2 + \frac{2}{m_3}(\Lambda_3+\Lambda_{12}+\Lambda_{23})}}
{1+\frac{\Lambda_{13}}{\Lambda_{31}}+\frac{\Lambda_{23}}{\Lambda_{32}}}
\right]. \
\eeqn
This would be the natural starting point for comparing the many-body and mean-field solutions of the
multiple-species
mixture,
for finite systems and at the infinite-particle-number limit.

\end{document}